\documentclass[a4paper]{cas-sc}
\usepackage[numbers]{natbib}
\usepackage{amssymb}
\usepackage{caption}
\usepackage{amsfonts}
\usepackage{amsmath}
\makeatletter
\newcommand*\rel@kern[1]{\kern#1\dimexpr\macc@kerna}
\newcommand{\widebar}{}
	\DeclareRobustCommand*\widebar[1]{ 
	\begingroup
	\def\mathaccent##1##2{%
		\rel@kern{0.8}%
		\overline{\rel@kern{-0.8}\macc@nucleus\rel@kern{0.2}}%
		\rel@kern{-0.2}%
	}%
	\macc@depth\@ne
	\let\math@bgroup\@empty \let\math@egroup\macc@set@skewchar
	\mathsurround\z@ \frozen@everymath{\mathgroup\macc@group\relax}%
	\macc@set@skewchar\relax
	\let\mathaccentV\macc@nested@a
	\macc@nested@a\relax111{#1}%
	\endgroup
}
\makeatother
\usepackage{fancyhdr}
\usepackage{lastpage}
\usepackage{booktabs}
\usepackage{xurl}
\usepackage{graphicx}
\usepackage{hyperref}
\usepackage{empheq}
\usepackage[title]{appendix}
\usepackage{rotating}
\usepackage{makecell}
\usepackage{diagbox}
\usepackage{longtable}
\usepackage{lscape}
\usepackage{subcaption}

\def\tsc#1{\csdef{#1}{\textsc{\lowercase{#1}}\xspace}}
\tsc{WGM}
\tsc{QE}

\begin{document}
	
	\shorttitle{Optimal Layout Plan via minimizing Electrostatic Energy}    
	
	\shortauthors{Ka Ian Im et. al.}  
	
	\title [mode = title]{Optimal Layout Plan of Stands at the Macao Food Festival via Minimizing the Electrostatic Potential Energy with the Effective Charge as Popularity of Stands}  
	
	
	
	%
	
	\author{Ka Ian Im} \author{In Kio Choi} \author{Pak Kio Lei} \author{Hou Fai Chan} \author{U In Ian} \author{Wei Shan Lee\corref{corr}}[type=editor,orcid=0000-0003-4801-0817
	]
	\cortext[corr]{Corresponding author: Wei Shan Lee}
	\ead{wslee@g.puiching.edu.mo}	
	\affiliation{organization={Pui Ching Middle School Macau},
		addressline={No.7, Avenida de Horta e Costa},
		city={Macau},
		country={China}}
	\begin{abstract}
		We proposed a mathematical model for designing the layout diagram of stand locations at the Macao Food Festival. The optimal layout diagram may be defined in such a way that, while requiring the distance between every pair of stands should not be too far away from each other, the crowd control is well managed so that people may patronize stands more effectively. More popular stands may have larger patronage, resulting in higher pedestrian flow nearby. Therefore, to avoid customers from packing shoulder to shoulder around more popular stands, we may treat every stand as a charged particle carrying an effective charge: the more popular a stand is, the higher the effective charge it carries. Under this assumption, the problem is then converted to the minimization problem of Coulomb electrostatic potential energy on a specific configuration of charge locations, with which the global minimum may be found by the Simulated Annealing and Metropolis Algorithm. Electrostatic energy density is interpreted as density of customers, while electric field the reversed crowd flow. Therefore, at a certain location we are able to predict the customer density by calculating the energy density and the net crowd flow with electric field lines. We also concluded that even though the required computation time to obtain a configuration of stand locations with the energy value close to the global minimum with a tolerable difference may be irrelevant to the randomly generated initial configuration of stand locations, setting up an appropriate initial configuration could be one of the key issues to find out the actual global minimum.
	\end{abstract}

	\begin{keywords}
		Optimization and Control \sep Macao Food Festival\sep Simulated Annealing and Metropolis Algorithm(SAMA)\sep Coulomb Electrostatic Potential Energy
	\end{keywords}
	
	\maketitle

	\section{Introduction}
	Macau is a special administrative region of China and it was once a colony of the Portuguese Empire. Under the mixing of Portuguese and Chinese culture, Macau has become a diversified city that preserves both cultural heritage like traditional food, historic buildings, etc. This unique cultural background attracts a large number of tourists and it led tourism to become one of the important economic income of Macau. Therefore, different types of traditional festival celebrations and national events are held. Known as the capital of gastronomy, Macao has a wide variety of cuisines and retains the local cultural flavor, Coexisting and blending with China and the West, so to meet the taste buds of tourists around the enjoyment. Macau was named as a Creative City of Gastronomy by UNESCO in 2017 because of its 400-year culinary heritage\cite{macao unesco cuisine}. 
	
	It is noted that local festivals are widely used as tourism promoters and boost regional economies\cite{Felsenstein and Fleischer}. Also, the local festival scene does fit the pattern of local sustainable economic development activities better than other scenes\cite{O'Sullivan and Jackson}. Because of the uniqueness of cuisine has become one of the important factors to attract tourists to Macau, some annual national events such as Macau Grand Prix and Macau Food Festival are held each November\cite{industry macao tourism}. Macau Food Festival which locates Sai Van Lake Square is one of the annual events that contains various culinary delights from Asia and Europe\cite{mtt macao tourism}. This landmark event gathers local and international chefs and key stakeholders from the sector from across Asia and Europe is a fine example of Macao's expertise in hosting large-scale gastronomy-related events\cite{macao unesco}. 
	
	Owing to the fact that the local cuisine is an important part of the travel experience\cite{Sengel and Karagoz}, Macau Food Festival takes food as an attraction to attract tourists to come specially to taste food and experience food culture\cite{Wang Xin}. Su\cite{Su WX} showed that the foreign tourists who flock to Macau's food festival prove that the government has come to see it as a tourist attraction.
	
	Since the utilization efficiency of regional tourism industry elements can be improved by a rational spatial organization of the tourism industry\cite{JL Hao et al}, various places are using optimal layout trying to optimize the facilities, such as proposing a macro layout planning method to better serve self-driving tour travelers\cite{Zhang XQ et al}, or layout optimization of tourist toilets\cite{Han L et al}. It can even be used to solve problems like using mathematical optimization to determine the optimal layout of a dining room during the era of COVID-19\cite{Contardo and Costa}, within the intended span of a metro station, address the layout planning of a public bicycle system\cite{JX Chen et al} or even improve the damping effect of the MR damper\cite{SP Hu et al}.
	
	However, no article to date has been done on optimal stands locations for the Macau Food Festival. Therefore, the purpose of this study was to optimize the booth layout of the Macau food Festival, with which 140 booths were allocated among 200 possible locations, by approaching through minimizing Coublumb electrostatic potential energy while treating the stand popularity as the effective charge (EffQ), which was acquired by questionnaire survey and data analysis, where we classified booths into more popular ones with higher values of EffQ and less popular ones with lower EffQs.

	\section{Methods}
	Suppose there are $\mathcal{N}$ possible locations for the stands at the Macao food festival heritage labeled as $[0,1,2,\dots,\mathcal{N}-1]$, and there are $N$ possible stands, where $N<\mathcal{N}$. Also, let $Q_{i}$ denote the effective charge (EffQ) of stand $i$. We then imagined a set of charged particles dispersing around in the 2-dimensional space. At the food festival, customers desire to patronize those more popular stands, equivalently referring to the highest charged particles. If the popular stands are located very near at the festival, it would be much easier for customers to get huddled together and also less convenient for customers to shop. To avoid crowding, our first idea was that it would be preferable to separate popular stands from one another as far as possible; nonetheless, it would cause inconvenience for customers to patronize popular stands. Keeping these in mind, we have to find out a layout plan whose stands are not too far away from one another while preventing most of the popular stands from locating close to one another. To the best of our imagination, this situation resembles that of the minimum energy of a system of two-dimensional electric charges. In other words, we would like to find out the locations for every charged particle (stand) such that the overall Coulomb electrostatic potential energy would be minimized. This is to say, we would like to find out the configuration of locations $\{ r_{i}, \forall i \}$, such that the overall electric potential energy
	\begin{center}
		\begin{equation}\label{summation}
			E\bigl(\{r_{i}, \forall i \}\bigr)=\sum_{i<j}\frac{Q_{i}Q_{j}}{|\vec{r_{i}}-\vec{r_{j}}|}
		\end{equation}
	\end{center}
	is minimized. We then tried to calculate the global minimum via the simulated annealing and Metropolis algorithm (SAMA)\cite{kirkpatrick}.

\begin{center}
		\begin{figure}[htbp]
			\centering
			\includegraphics[width=1.0\textwidth]{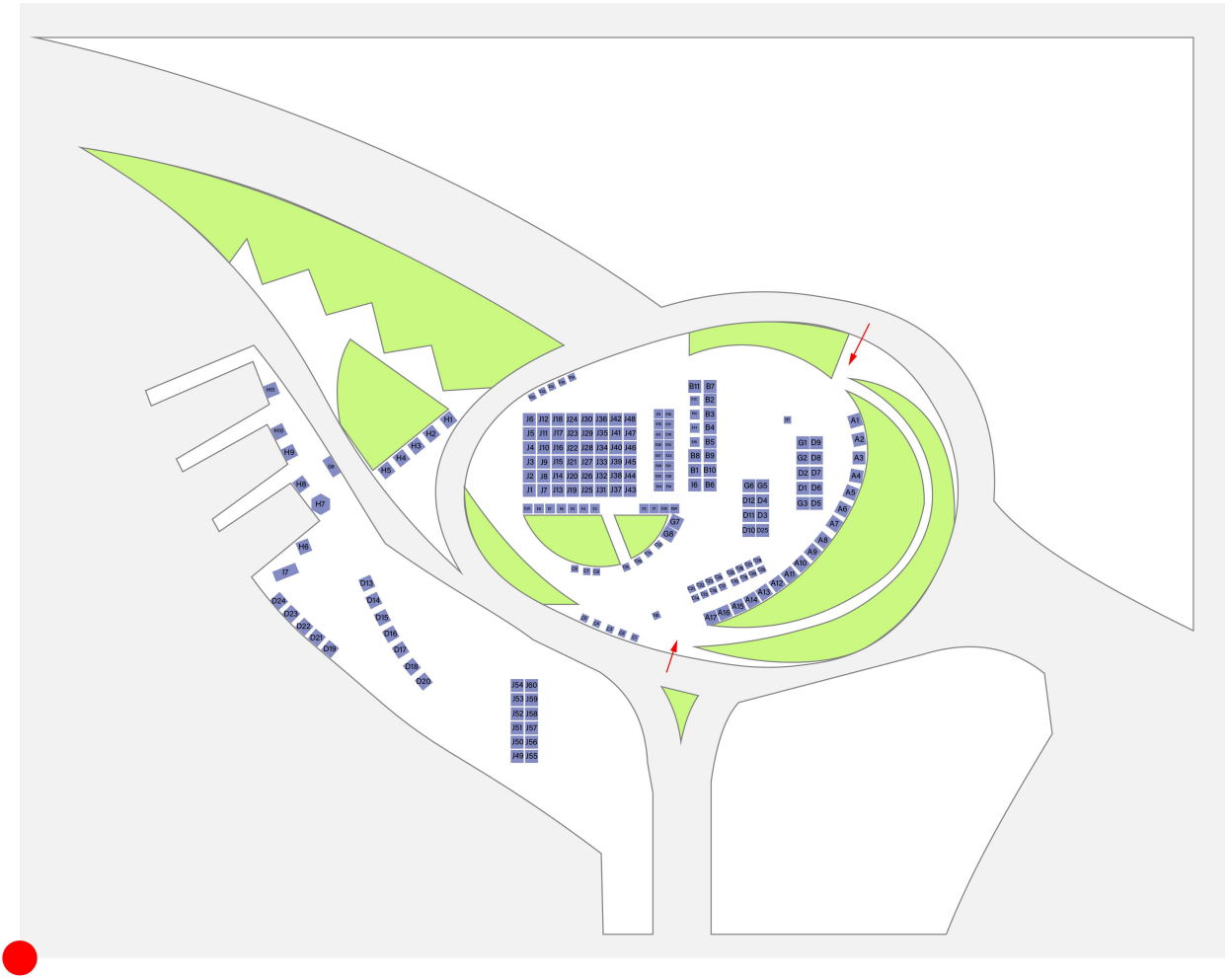}
			\caption{Locations of $140$ stands at the Macau Food Festival and $60$ other empty locations. Details for the stands numbers and coordinates were listed in Table $\ref{tab:Names, Coordinates and ID numbers of Stands}$}
			\label{fig:coordinates}
		\end{figure}
	\end{center}

	The idea of the algorithm may be summarized as follows. While the probability for the system with energy $E_{i}$ is described by the Boltzmann distribution with $\beta=1/\tau=1/k_{B}T$, $k_{B}$ being the Boltzmann constant:
	\begin{center}
		\begin{equation}\label{Boltzmann} 
			P(E_{i})\propto e^{-\beta E_{i}},
		\end{equation}
	\end{center}	
	we arbitrarily generate an initial configuration of locations of our stands, then calculate the initial value of the quantity $E_{i}$ in Eq.($\ref{summation}$). Then, we generate another trial configuration of locations of stands by randomly picking up two locations of stands to swap, calculating again the quantity after the change in Eq.($\ref{summation}$), say $E_{j}$. If the new quantity is smaller than the old one, we accept the swap. However, if the new quantity is larger than the old one, we may still try to accept the new configuration by introducing the Metropolis Algorithm with the following criteria\cite{Landau Paez and Bordeianu},\cite{Krauth}:
	\begin{center}
		\begin{equation}\label{Metropolis probability} 
			P_{\beta}(E_{j})=\left\{
			\begin{array}{ll}
				\hspace{20pt}1\hspace{35pt}\mathrm{if\hspace{5pt}}  E_{j}\leq E_{i}; \\
				e^{-\beta (E_{j}-E_{i})}\hspace{15pt}\mathrm{if\hspace{5pt}} E_{j}>E_{i}.\\
			\end{array}
			\right.
		\end{equation}
	\end{center}
	In practice, we generated a random number $z\in [0,1]$, then if $z$ satisfies 
	\begin{center}
		\begin{equation}\label{random number z} 
			z\leq e^{-\beta (E_{j}-E{i})},
		\end{equation}
	\end{center}
	we accept the swap; otherwise we reject the swap and go back to the previous configuration of stand locations. Throughout our study we have chosen values of the maximum temperature $\mathrm{T_{max}}$, minimum temperature $\mathrm{T_{min}}$, and $\mathrm{\tau=k_{b}T}$ as listed in Table $\ref{tab:parameters}$. It is interesting to indicate that Karabin and Stuart\cite{Karabin and Stuart} demonstrated that one may improve the efficiency of classical SAMA by allowing the temperature-varying $\mathrm{k_{b}(T)}$.
	\begin{center}
		\begin{table}[htbp]
			\centering
			\begin{tabular}{ccc}
				\hline\hline
				$\mathrm{T_{max}}$ & $\mathrm{T_{min}}$  & $\mathrm{\tau=k_{b}T}$ \\\hline
				$1.0$              & $1\times 10^{-2}$   & $1\times 10^{4}$       \\
				\hline\hline
			\end{tabular}
			\caption{Parameters used in SAMA.}
			\label{tab:parameters}
		\end{table}
	\end{center}
	Figure $\ref{fig:coordinates}$ indicates the map of the event. It shows the locations of all the $140$ booths that were designated in the year 2020, together with $60$ more empty locations we deliberately chose to allow possibilities for the stands to select. Each booth has a different ID number on it. Remade and redrawn were made on the original map acquired from the official Facebook account of Macau Food festival,\cite{official FB account} and stands were divided into eight types\cite{FB 18 categories}. The red dot in Figure $\ref{fig:coordinates}$ represents the origin of the coordinate so we could get the coordinates of the booths in Table $\ref{tab:Names, Coordinates and ID numbers of Stands}$. In addition, the arrows stand for the entrance of the event.\\
		
	In order to determine the value of $Q_{i}$ of a stand in Eq.($\ref{summation}$), we conducted a questionnaire survey. The first idea was that it would be better off if we may ask respondents to evaluate all 140 stands. However, it is not practical because it is time consuming for a person to answer. Therefore, we divided the questionnaire into eight parts. Besides a questionnaire of all \textit{types} of stands, which we called \textit{Grand} questionnaire as shown in Table $\ref{tab:grand questionnaire}$, for every type of stands, we also made a questionnaire of the subcategory consisting of all stands for that particular type. The respondents may choose to answer either all of the eight questionnaires or only one of them. The popularity of every stand in the particular subcategory was denoted as average (together with the standard deviation) in the following tables: Chinese restaurant zone (Table $\ref{tab:Chinese restaurant zone}$), European delicacies zone(Table $\ref{tab:European delicacies zone}$), Dessert zone(Table $\ref{tab:Dessert zone}$), Asia delicicas zone(Table $\ref{tab:Asia delicicas zone}$), Local delicacies zone (Table $\ref{tab:Local delicicas zone}$), Sponsor and other(Table $\ref{tab:Sponsor and other}$), and Game booth(Table $\ref{tab:Game booth}$). Notice that the Type Sponsor and Type Other were grouped together in the survey and results were listed together in Table $\ref{tab:Sponsor and other}$. Afterward, the effective charge $Q_{i}$ of each stand was obtained by multiplying the value of the type in the grand questionnaire with the value of that particular stand in the subcategory. \\
	The error propagation was also considered. Suppose that there are two mean values $\widebar{x}$ and $\widebar{y}$ from the grand questionnaire and one of the subcategories, with the standard deviation of mean $\sigma_{\widebar{x}}$ and $\sigma_{\widebar{y}}$, respectively. The $Q_{i}$ of the stand may then be understood as $Q_{i}=\widebar{xy}=\widebar{x}\cdot\widebar{y}$, and the standard deviation of mean for $\widebar{xy}$ is\cite{error propagation}
	\begin{center}
		\begin{equation}\label{error propagatoin}
		\sigma_{\widebar{xy}}=\sqrt{		
		\bigg(  \frac{  \sigma_{\widebar{x}} }{\widebar{x}}    \bigg)^2	
		+
		\bigg(  \frac{  \sigma_{\widebar{y}} }{\widebar{y}}    \bigg)^2
		}\times \widebar{xy}.
		\end{equation}
	\end{center}

	After acquiring all the $Q_{i}$s, we normalized them with 
	\begin{center}
		\begin{equation}\label{standarScaler}
	Q\prime=\frac{Q-\widebar{Q}}{\sigma},
		\end{equation}
	\end{center}
	where $\widebar{Q}$ is the mean of the effective charge and $\sigma$ is the standard deviation for $Q$. By doing so, the $Q\prime$s have properties that mean $\widebar{Q}\prime=0$ and variance is equal to 1\cite{standardScaler}. 
	As a result, the normalized effective charges $Q\prime$ of all stands were listed in Table $\ref{tab:norm Q}$.
	\section{Results and Discussions}
\begin{center}
	\begin{figure}[!htbp]
		\centering
		\includegraphics[width=0.75\textwidth]{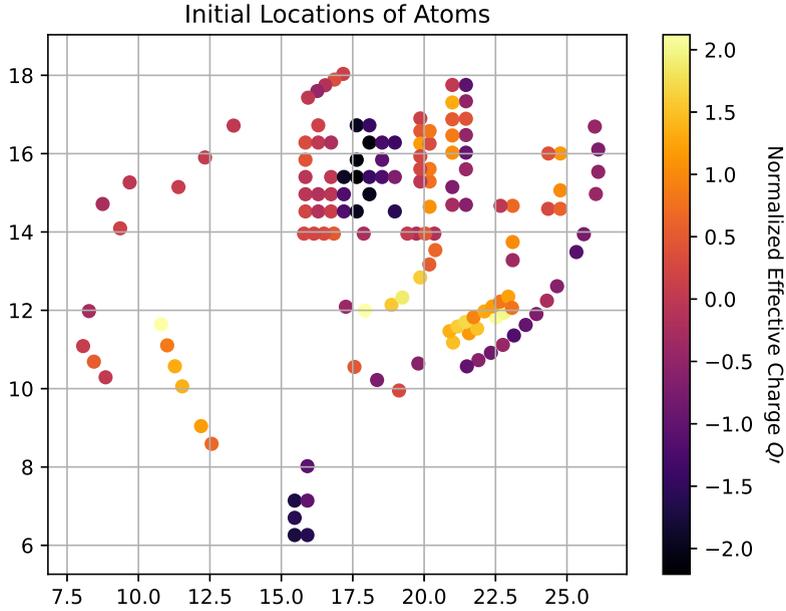}
		\caption{At the beginning of algorithm, $140$ stands were randomly deployed among $200$ possible locations. $60$ more empty locations were also allowed to be selected for swapping. }
		\label{fig:initLocsOfAtoms}
	\end{figure}
\end{center}

\begin{center}
	\begin{figure}[!htbp]
		\centering
		\includegraphics[width=1.0\textwidth]{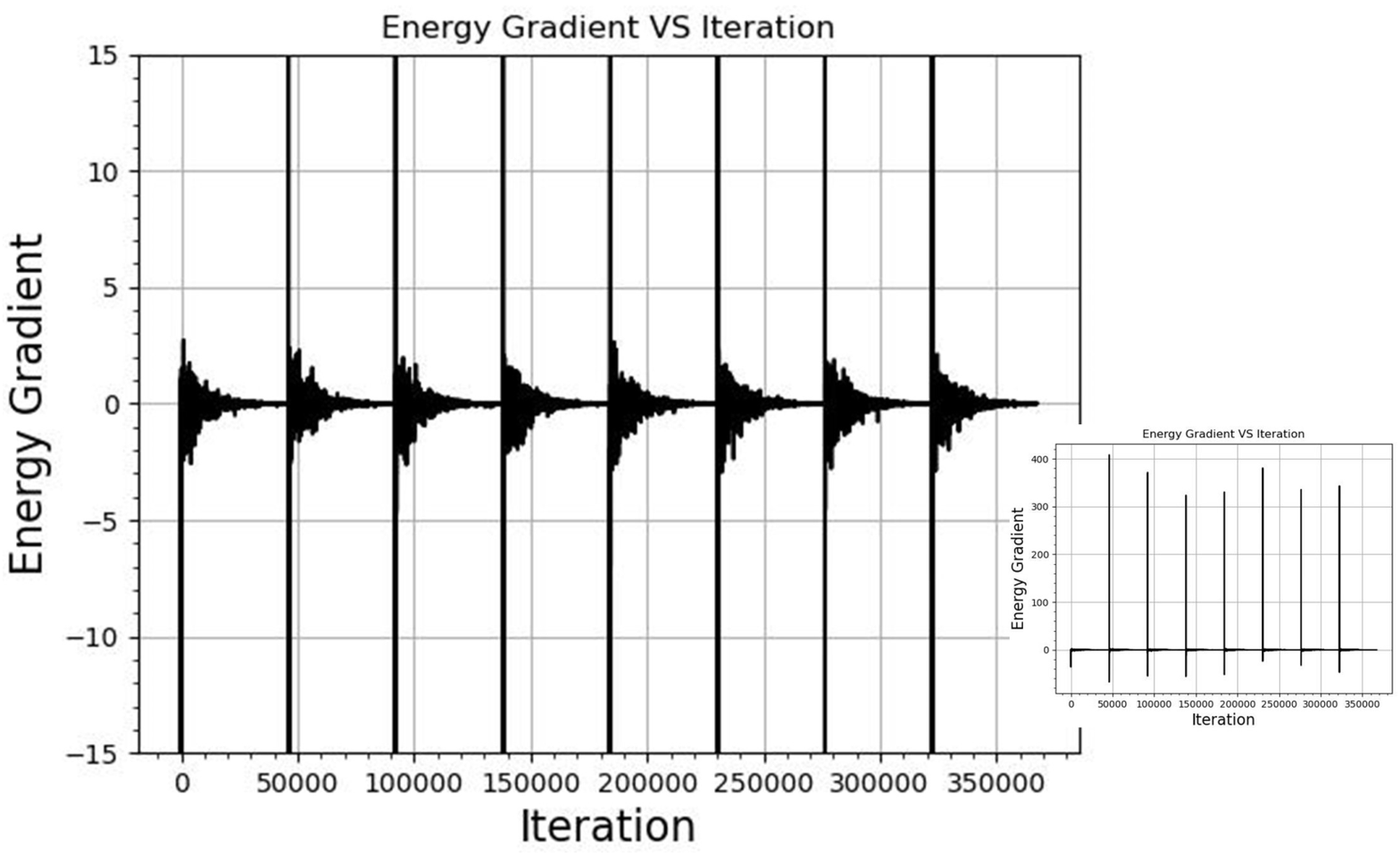}
		\caption{Energy gradient vs. Iteration. Notice that the shapes of envelops within the several periods were also similar, which probably means that computation time required to search for the configuration with an acceptable energy value is not dependent on the initial configuration of our system. (Inset) The overview of energy gradient vs. Iteration. Shark peaks indicated the reset of initial conditions of booth locations.}
		\label{fig:energyGradient}
	\end{figure}
\end{center}

\begin{center}
	\begin{figure}[!htbp]
		\centering
		\includegraphics[width=0.75\textwidth]{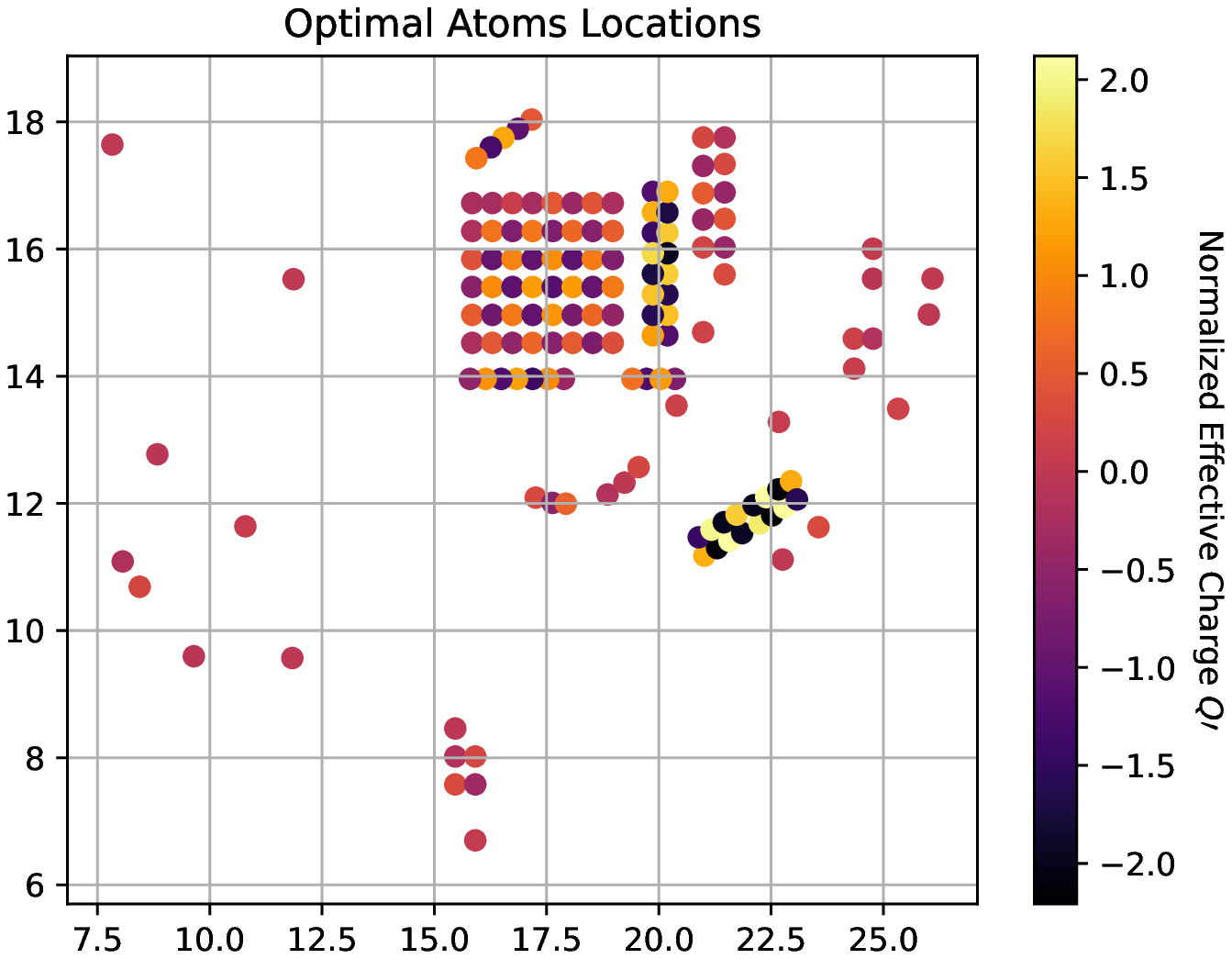}
		\caption{Optimal locations of stands, together with values of $Q\prime$ as various colors indicated at the color bar.}
		\label{fig:optimalLocOfAtoms}
	\end{figure}
\end{center}

\begin{center}
	\begin{figure}[!htbp]
		\centering
		\includegraphics[width=0.8\textwidth]{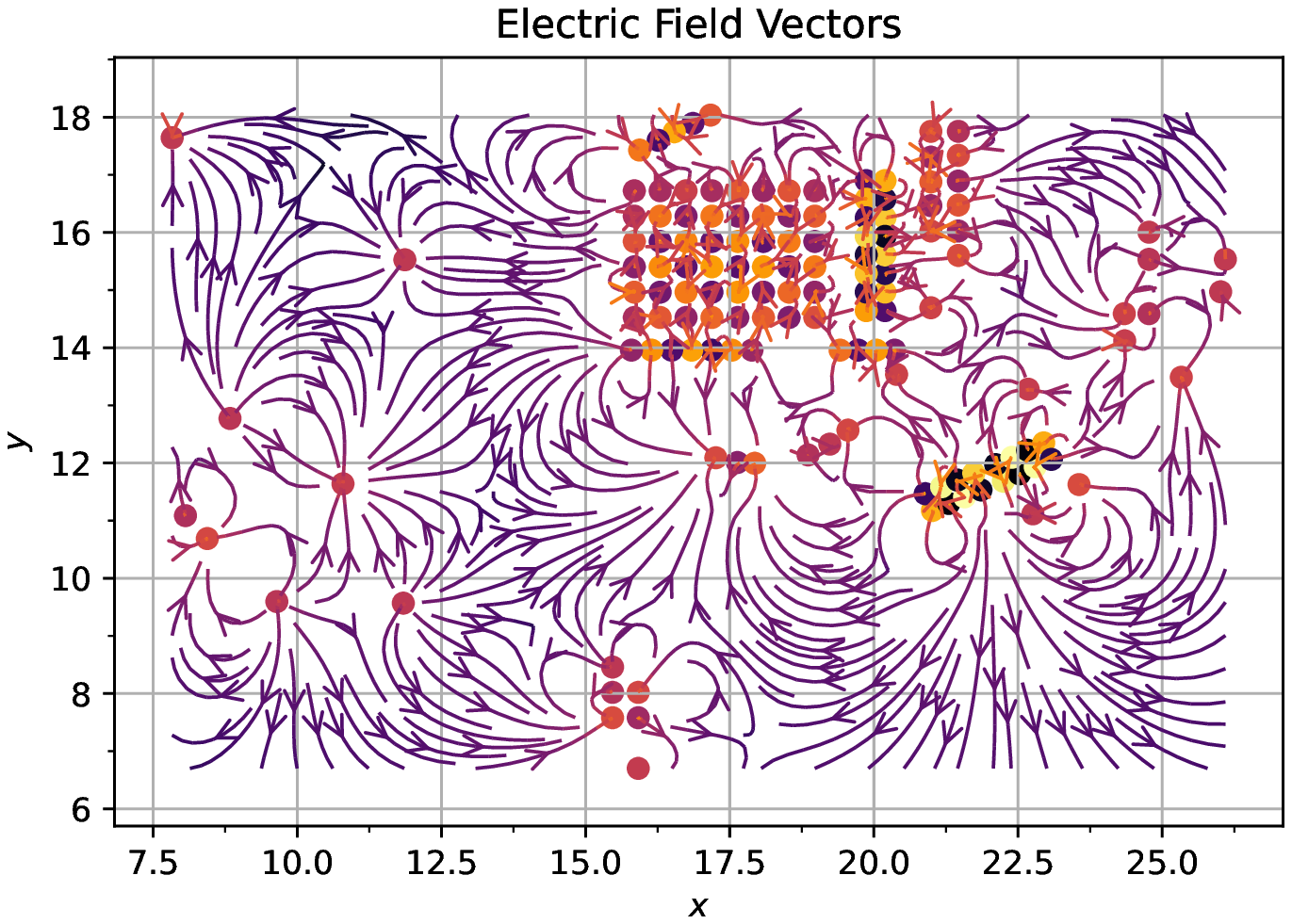}
		\caption{Electric field vectors for the system of booths with $Q\prime$, which could help us predict the net crowd flow at a specific location. }
		\label{fig:electricFieldLine}
	\end{figure}
\end{center}

\begin{center}
	\begin{figure}[!htbp]
		\centering
		\includegraphics[width=0.8\textwidth]{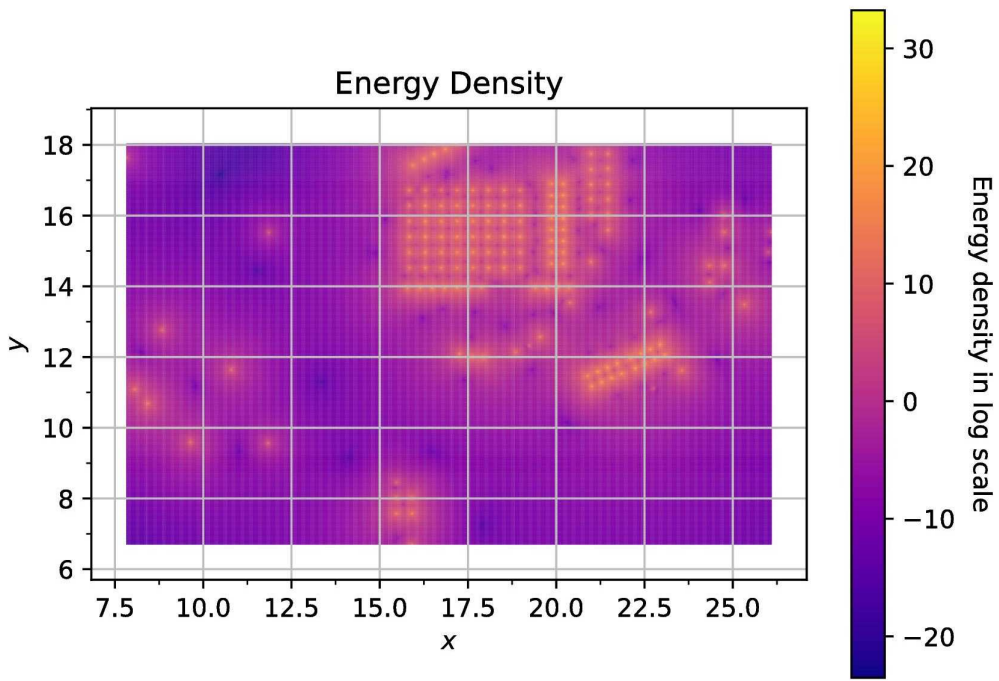}
		\caption{Logarithmic energy density of the optimal layout, from which the customer density was able to predict.}
		\label{fig:energyDensity}
	\end{figure}
\end{center}

Figure $\ref{fig:initLocsOfAtoms}$ illustrated the initial locations of stands which was generated randomly at the very first. The color bar at the right-hand side indicated the normalized effective charge $Q\prime$ mentioned in Eq.($\ref{standarScaler}$) with brighter color indicating higher $Q\prime$. As the figure showed, most of the atoms (stands) were focused on the area of the y-axis $14$ to $18$, and the x-axis $15.0$ to $22.5$. As we could see around the two-column booths at $x = 20.0$, the stands were quite close to each other. In spite of this, there was also a diffuse community: dispersing in the area of the x-axis from $7.5$ to $15.0$, and the y-axis $8$ to $18$. In addition, some booths formed an arch shape at the right-side of the diagram. The initial energy calculated by Eq.($\ref{summation}$) was $477.55282$. The complete list of stand locations were given in Table $\ref{tab:coordinates of initial layout plan}$.
After randomly generating the above initial configuration of locations, we implemented the SAMA algorithm to try to find out the optimal value of energy by randomly picking up two locations to swap. In order to find out lower energy with some configuration of space, we allocated $60$ more locations besides the original $140$ ones which were already occupied by the stands. Under this circumstance, there were three possible cases when we randomly chose two locations to swap. The first case was that two of the locations were occupied by one booth each, therefore there was no problem for us to swap between them. Secondly, it could also be possible that only one of the two locations we picked up was occupied with a booth while the other was empty. In this case, we could still swap the two locations by carefully identifying the empty location. One more possibility that should be avoided was that we may also picked up two originally empty locations, which did not make any sense to swap between them.\\
\begin{center}
	\begin{figure}[!htbp]
		\centering
		\includegraphics[width=0.75\textwidth]{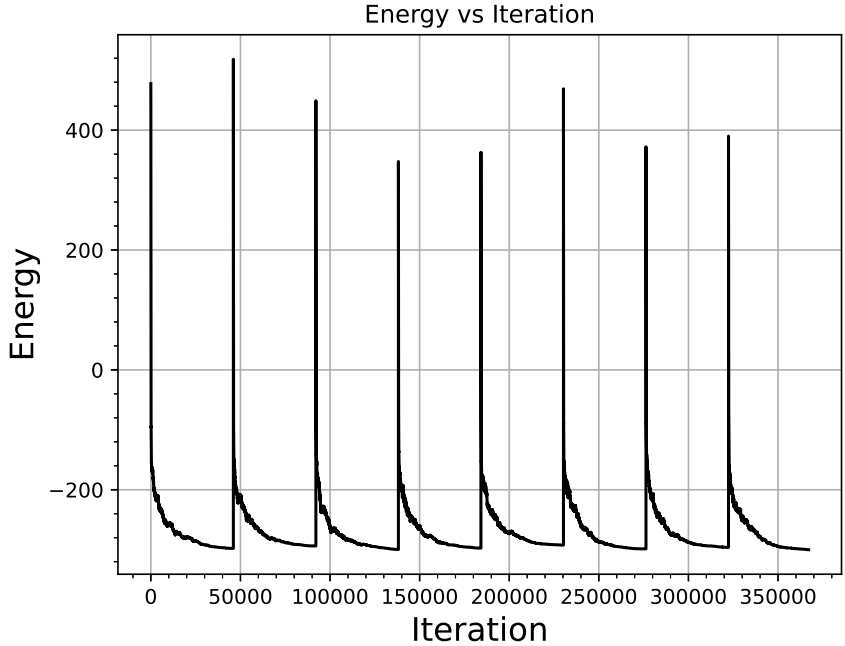}
		\caption{Energy vs. Iteration in the algorithm. The whole process may be roughly divided into eight periods along the Iteration (with the gap of 46 iterations), each starting up with a randomly generating initial configuration of stand locations, reducing down to energy around $-299$. The algorithm was set up to stop when it found an energy lower than $-300$.}
		\label{fig:energyVsIteration}
	\end{figure}
\end{center}
Figure $\ref{fig:energyVsIteration}$ illustrated energy vs. iteration, which may be divided into eight periods along Iteration axis, with the interval of roughly $46$ thousand iterations each. The first period started from the initial energy around $440$. In the process of running the SAMA, it dropped very quickly from the initial value down to $-200$ within less than $2000$ iterations, whereas it took $46$ thousand more iterations to reduce energy down to around $-299$ with some shapes of serration, with the layout shown in Fig $\ref{fig:animation2}$. As we discussed earlier, this configuration may not be the global minimum. Therefore after $46$ thousand iterations we randomly set up another initial configuration and started over the algorithm again, as shown in Fig $\ref{fig:animation3}$.\\
In the second period within Iteration $46$ thousand to $92$ thousand, energy dropped from $520$ to $-294$ (see also Fig $\ref{fig:animation3}$ and Fig $\ref{fig:animation4}$) before we set up another initial condition as in Fig $\ref{fig:animation5}$. Similar patterns occurred in the following consecutive five periods, as indicated through Fig $\ref{fig:animation5}$ to Fig $\ref{fig:animation15}$. However, since SAMA was purportedly designed to continue running until it found out an energy lower than $-300$, it did not succeed until in the eighth period. We may see that even if figures at the left-hand sides of Fig.$\ref{fig:animations}$ were all very close to $-300$, for the SAMA to reach the optimal (or even more importantly, the global) minimum, it could be crucial for the system to start at the appropriate initial configuration, as indicated in Fig.$\ref{fig:animation15}$. Snapshots in Fig.$\ref{fig:animations}$ were taken from the animation videos\cite{animation} of stands locations in company with the aforementioned featuring iterations in Figure $\ref{fig:energyVsIteration}$. 
\begin{center}
	\begin{figure}[!htbp]
		\begin{subfigure}{.45\textwidth}
			\centering
			\includegraphics[width=1.0\linewidth]{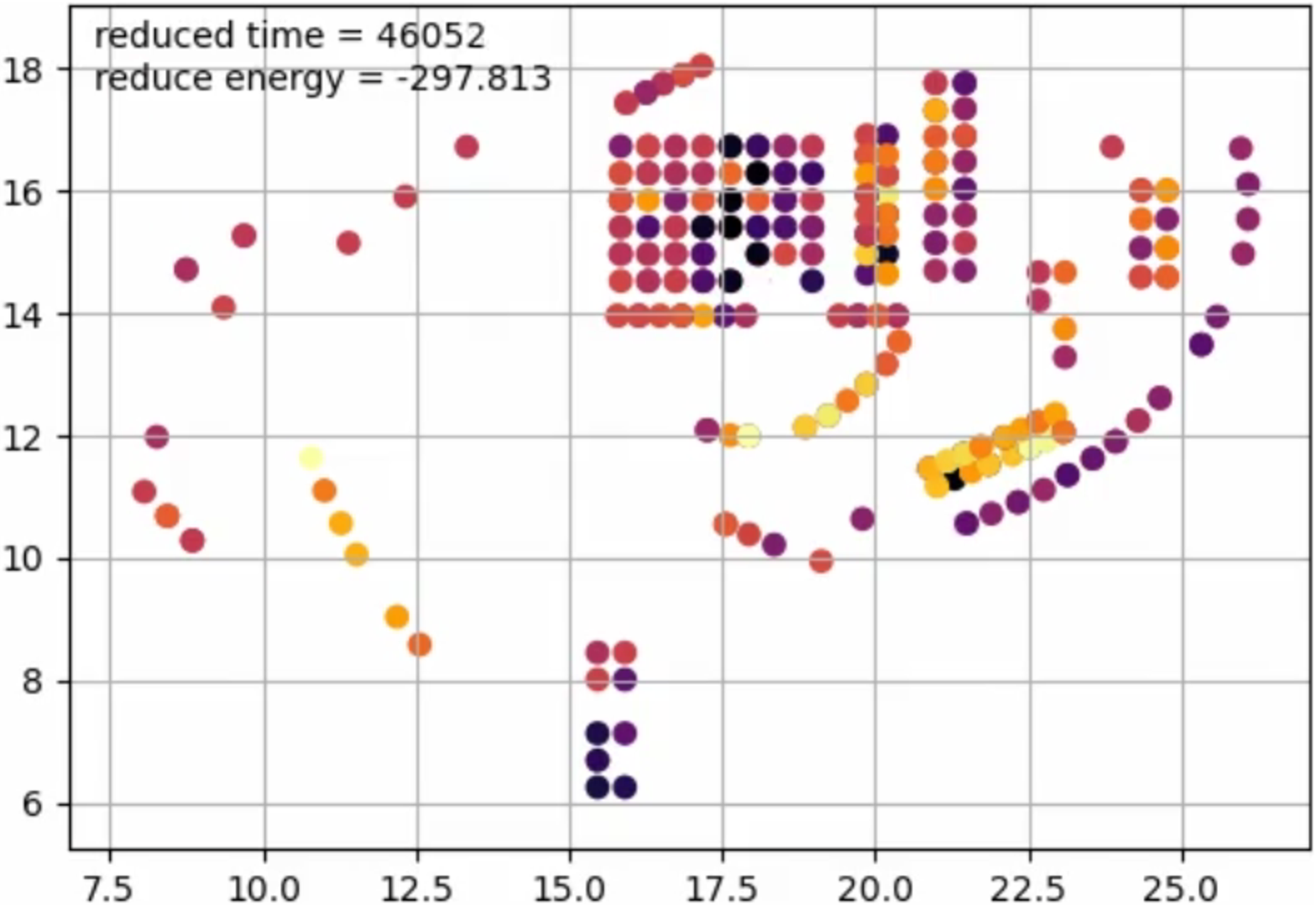}
			\caption{}
			\label{fig:animation2}
		\end{subfigure}
		\begin{subfigure}{.45\textwidth}
			\centering
			\includegraphics[width=1.0\linewidth]{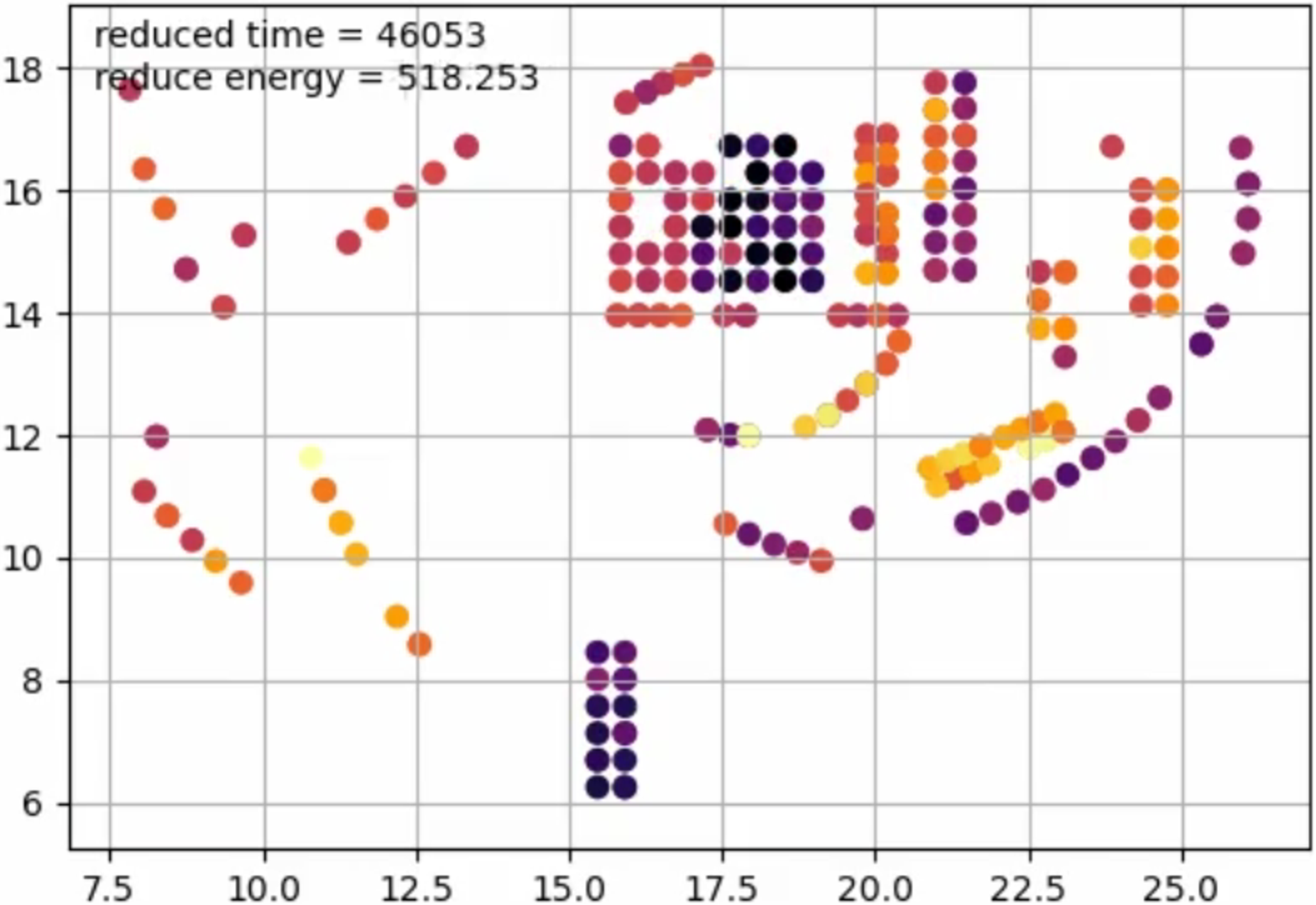}
			\caption{}
			\label{fig:animation3}
		\end{subfigure}
		\begin{subfigure}{.45\textwidth}
			\centering
			\includegraphics[width=1.0\linewidth]{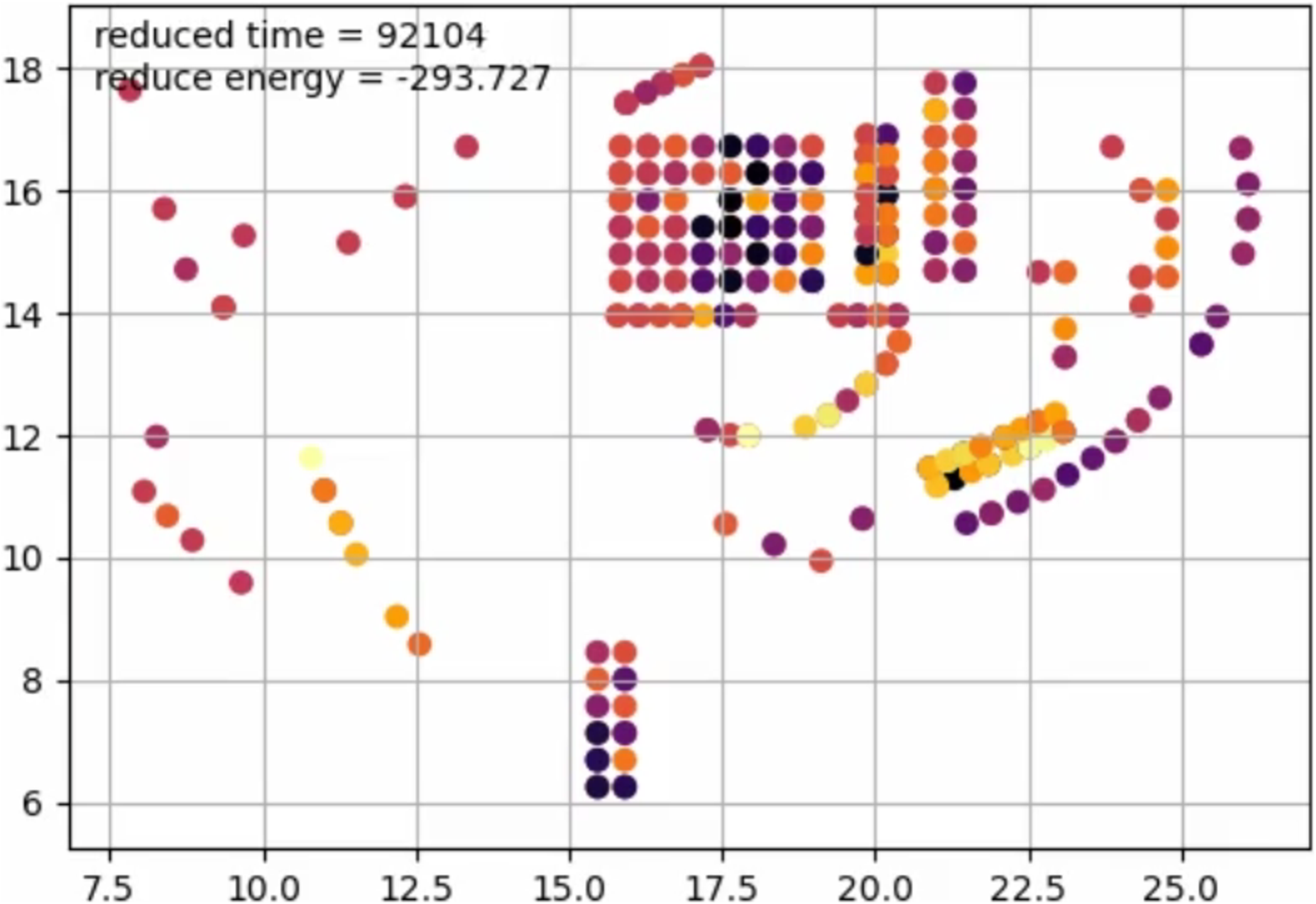}
			\caption{}
			\label{fig:animation4}
		\end{subfigure}
		\begin{subfigure}{.45\textwidth}
			\centering
			\includegraphics[width=1.0\linewidth]{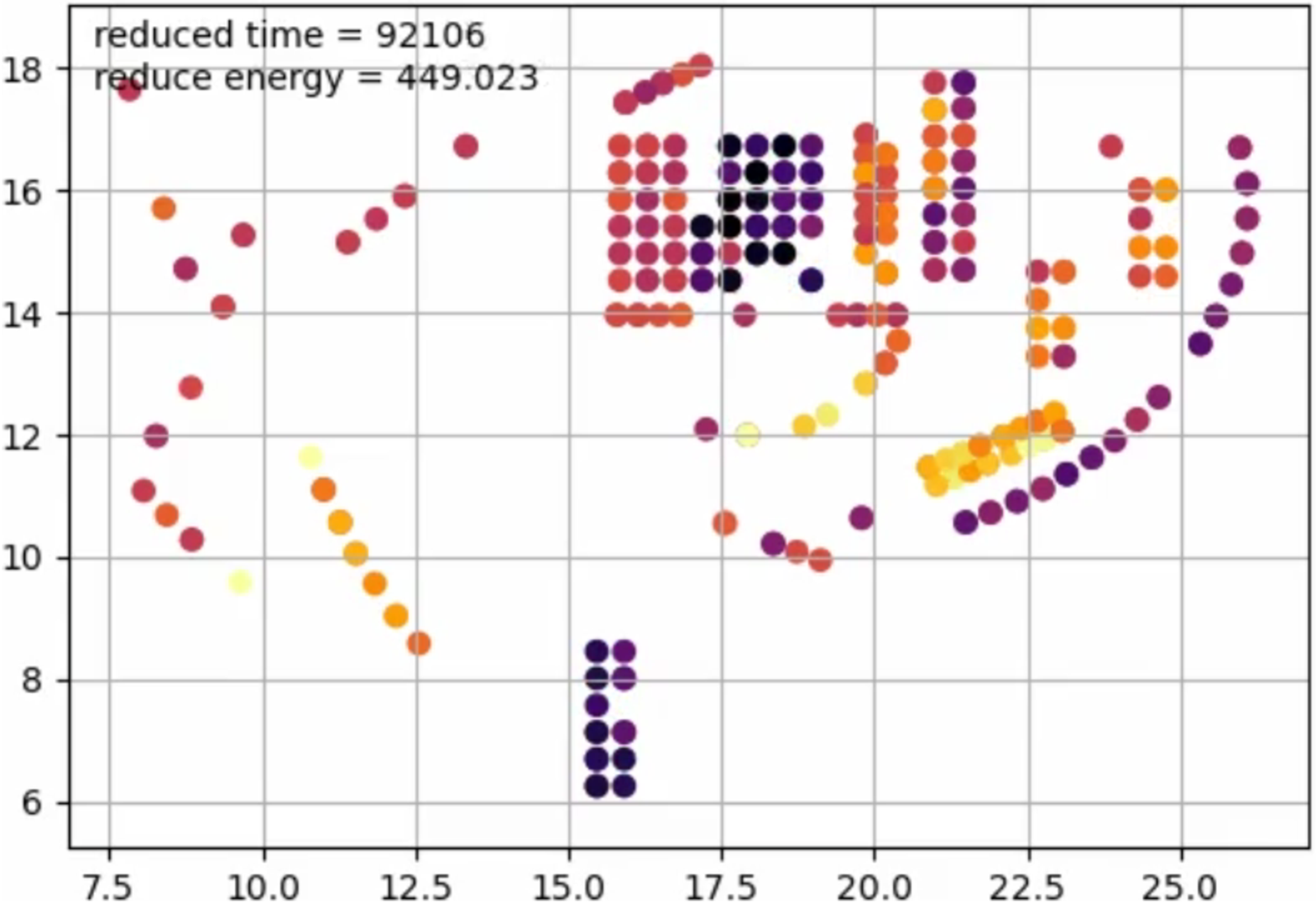}
			\caption{}
			\label{fig:animation5}
		\end{subfigure}
		\begin{subfigure}{.45\textwidth}
			\centering
			\includegraphics[width=1.0\linewidth]{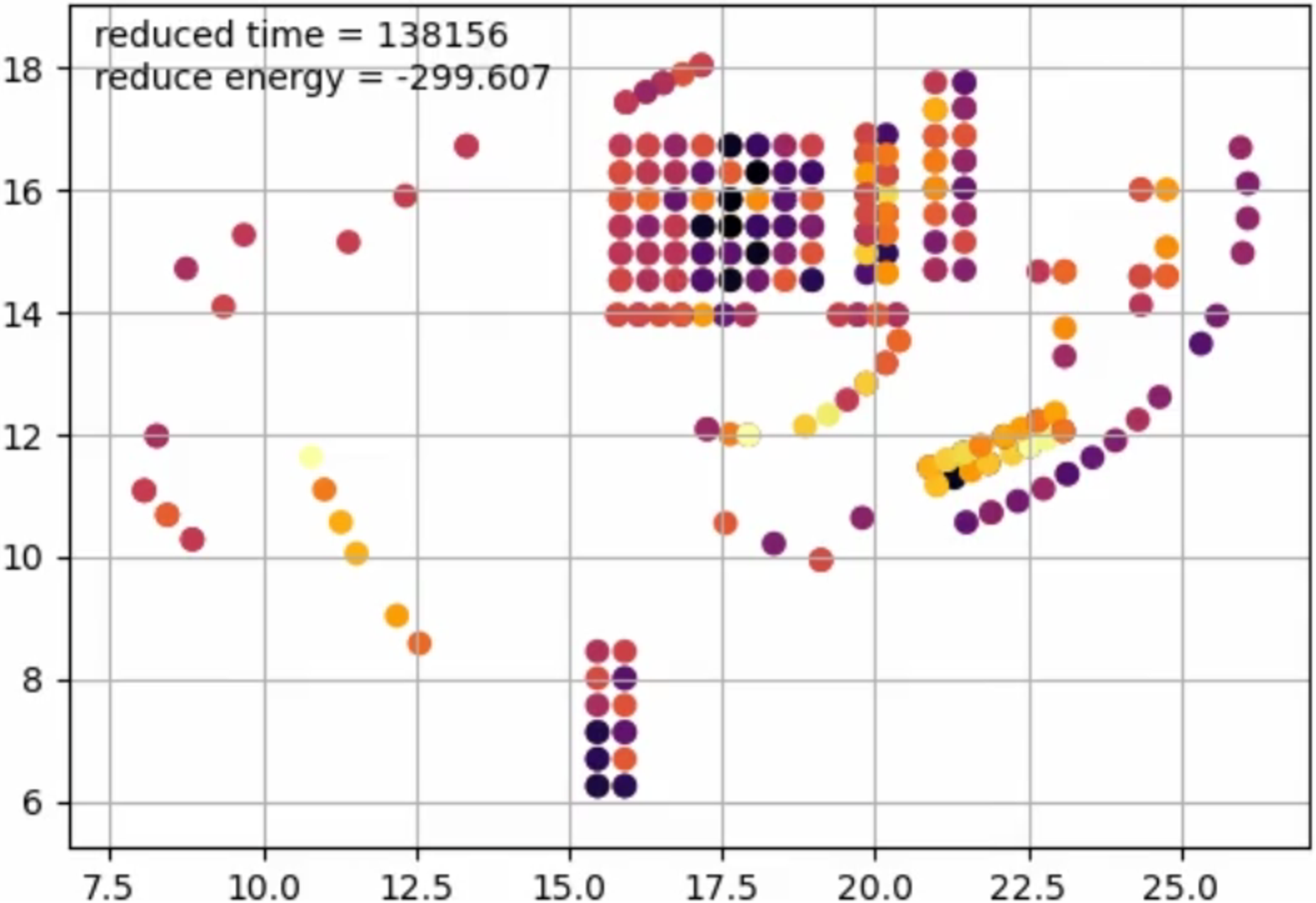}
			\caption{}
			\label{fig:animation6}
		\end{subfigure}
		\begin{subfigure}{.45\textwidth}
			\centering
			\includegraphics[width=1.0\linewidth]{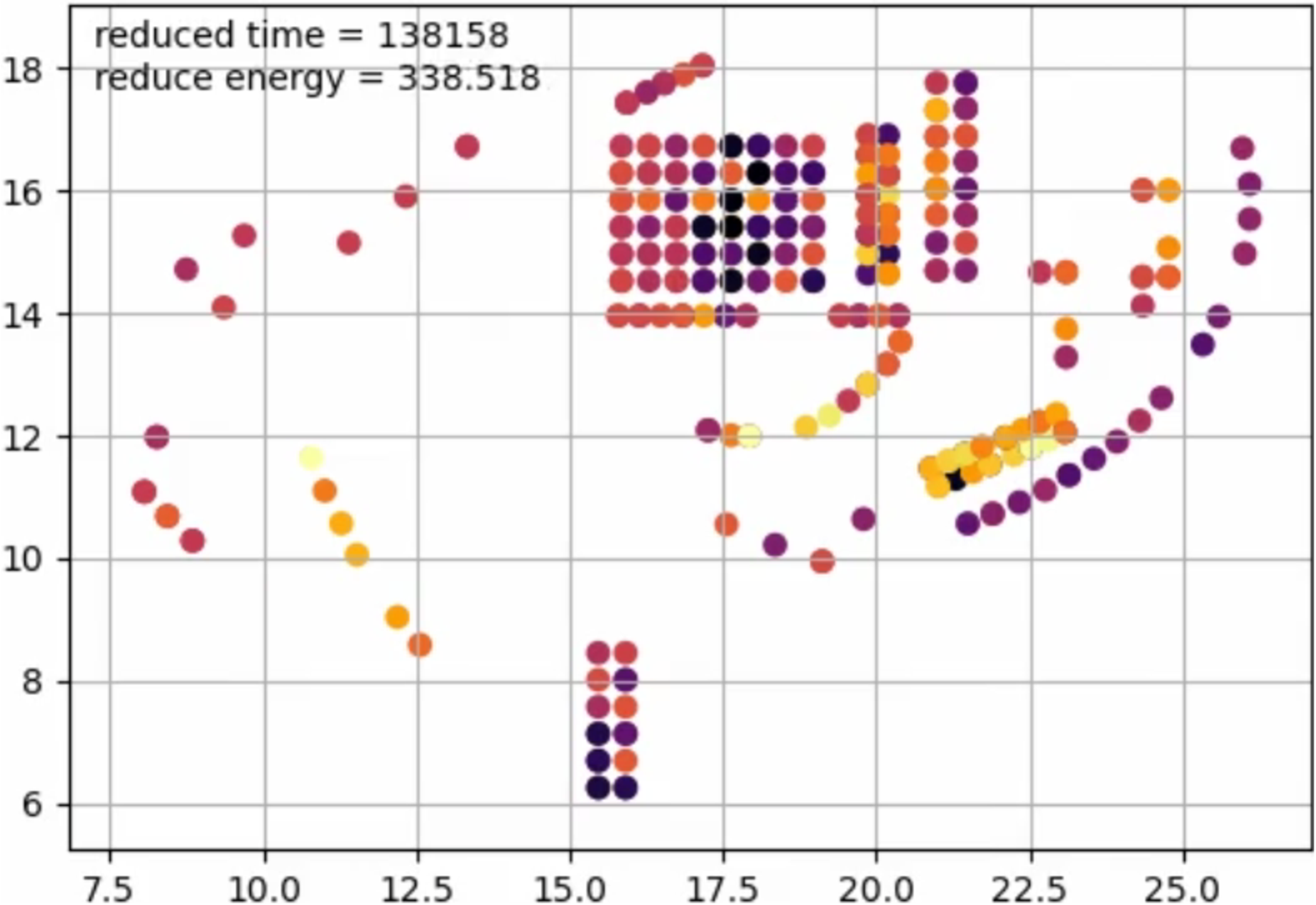}
			\caption{}
			\label{fig:animation7}
		\end{subfigure}
		\begin{subfigure}{.45\textwidth}
			\centering
			\includegraphics[width=1.0\linewidth]{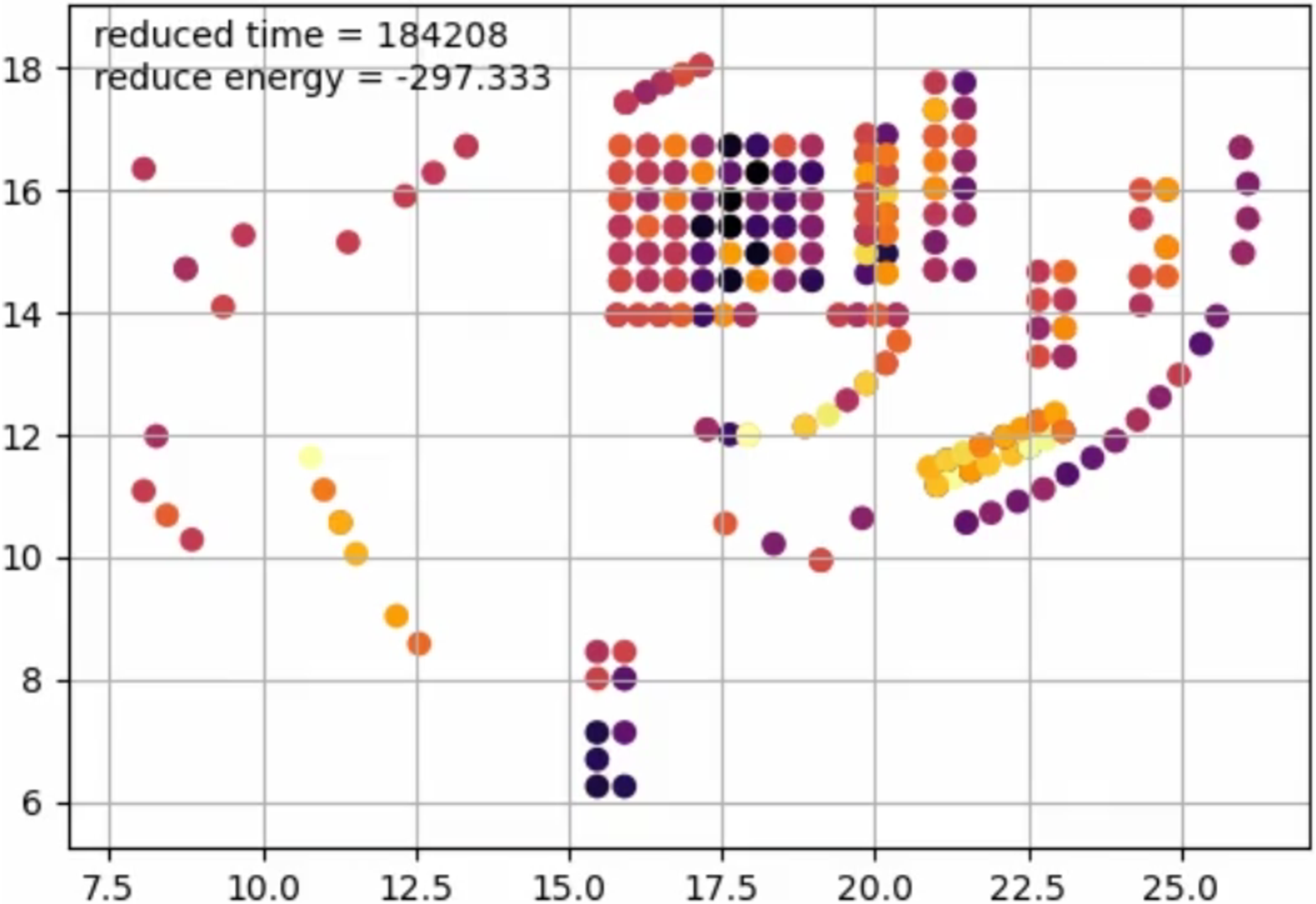}
			\caption{}
			\label{fig:animation8}
		\end{subfigure}
		\begin{subfigure}{.45\textwidth}
			\centering
			\includegraphics[width=1.0\linewidth]{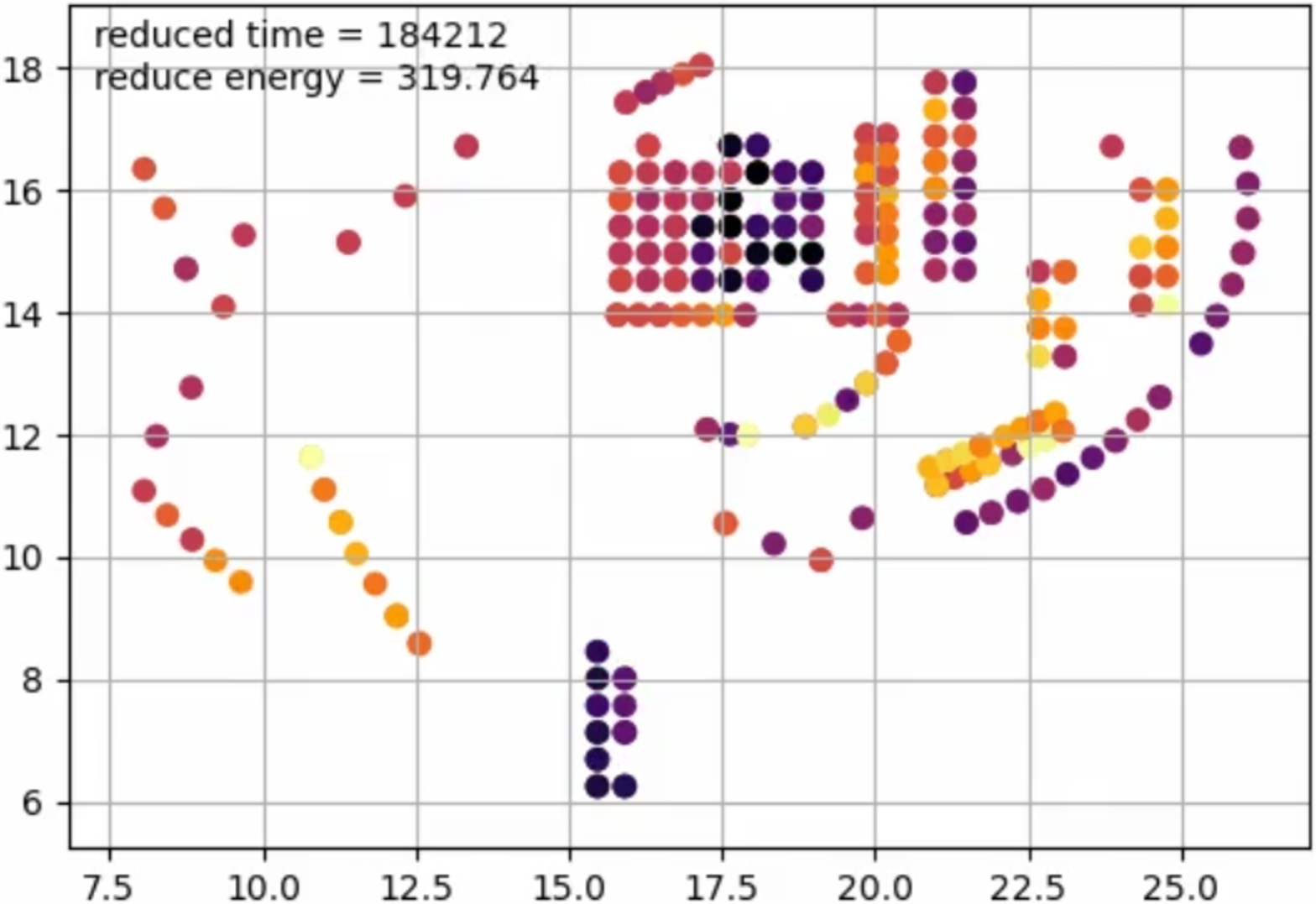}
			\caption{}
			\label{fig:animation9}
		\end{subfigure}
	\end{figure}
\end{center}
\begin{center}
	\begin{figure}\ContinuedFloat
\begin{subfigure}{.45\textwidth}
	\centering
	\includegraphics[width=1.0\linewidth]{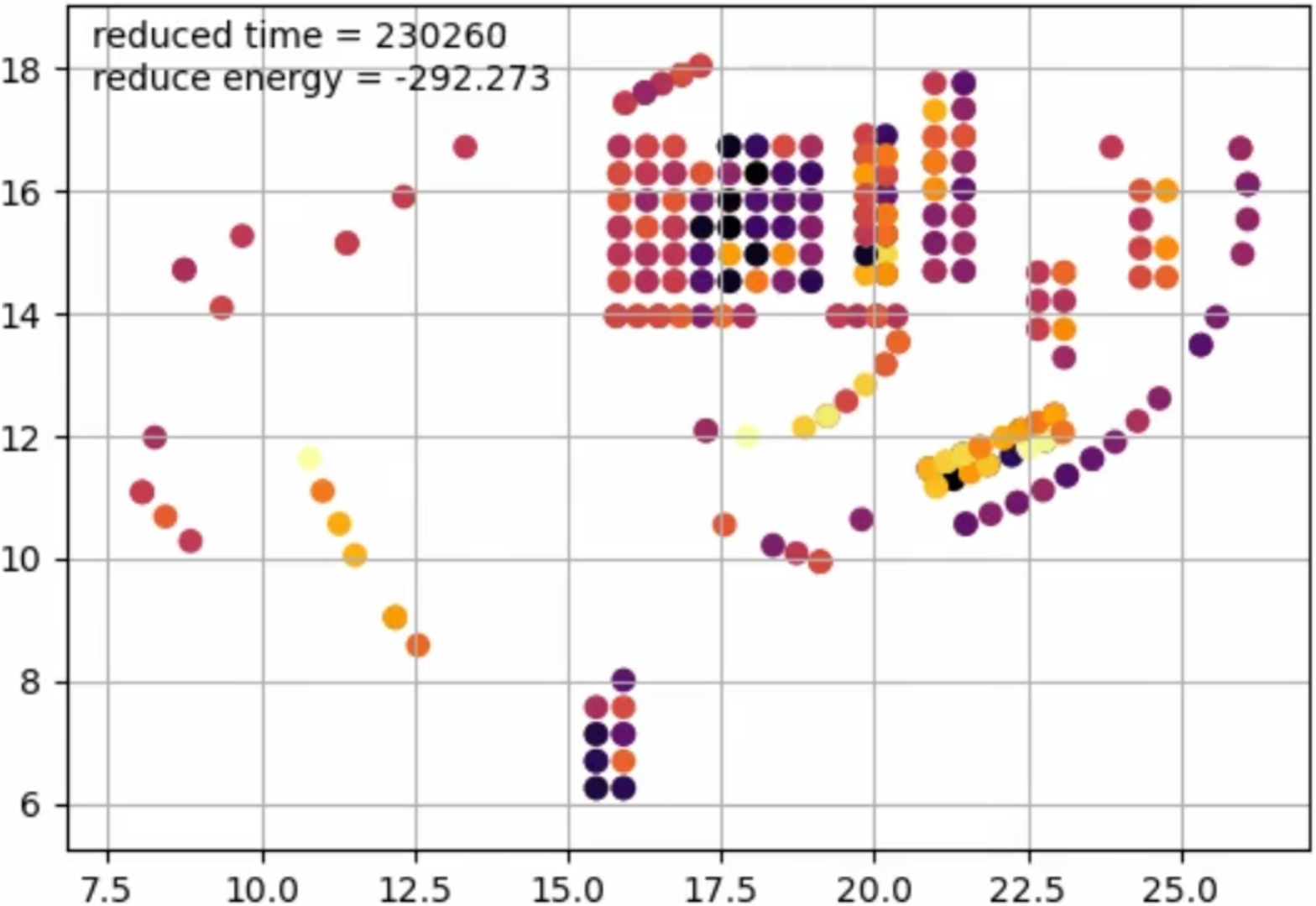}
	\caption{}
	\label{fig:animation10}
\end{subfigure}
\begin{subfigure}{.45\textwidth}
	\centering
	\includegraphics[width=1.0\linewidth]{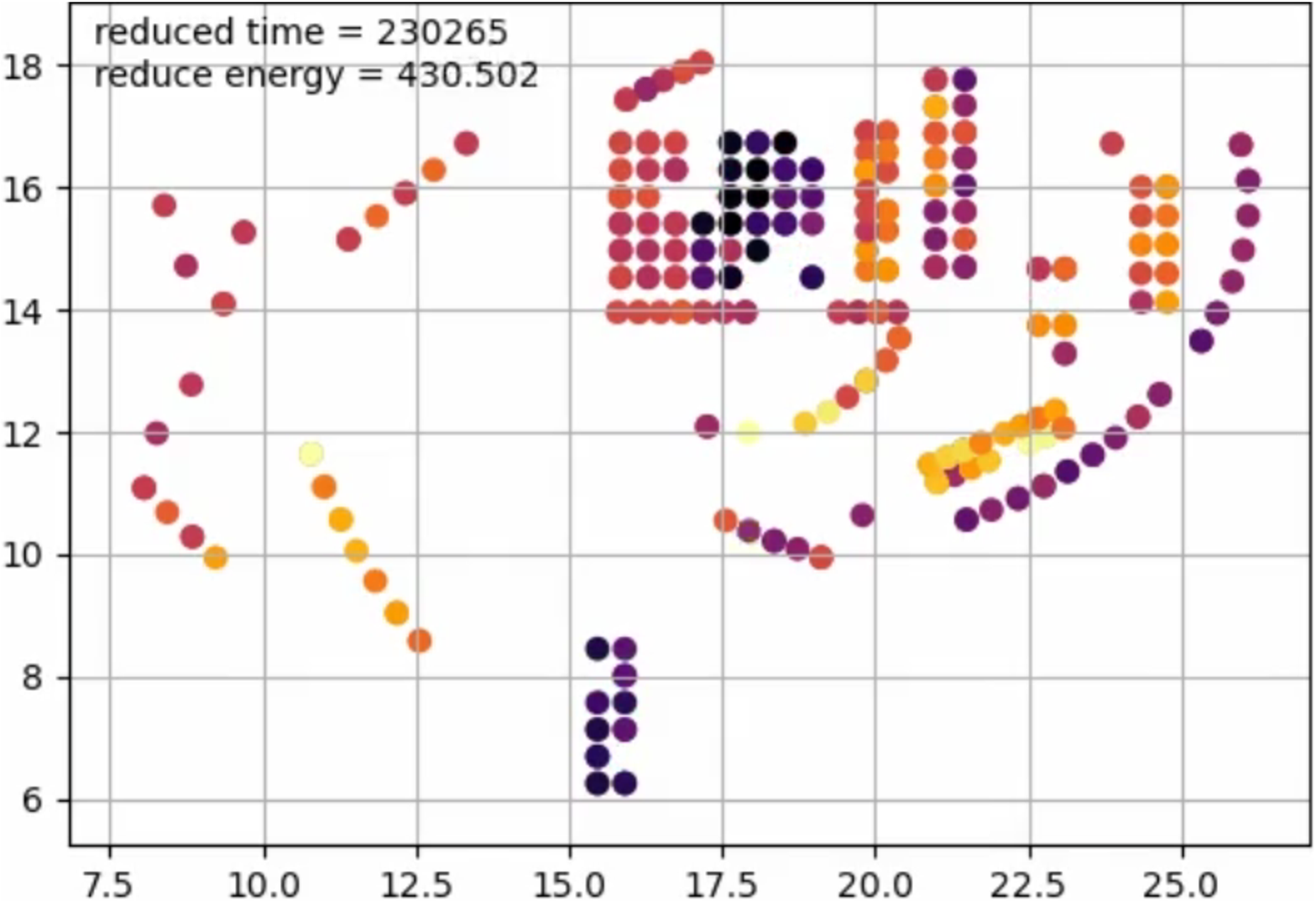}
	\caption{}
	\label{fig:animation11}
\end{subfigure}
\begin{subfigure}{.45\textwidth}
	\centering
	\includegraphics[width=1.0\linewidth]{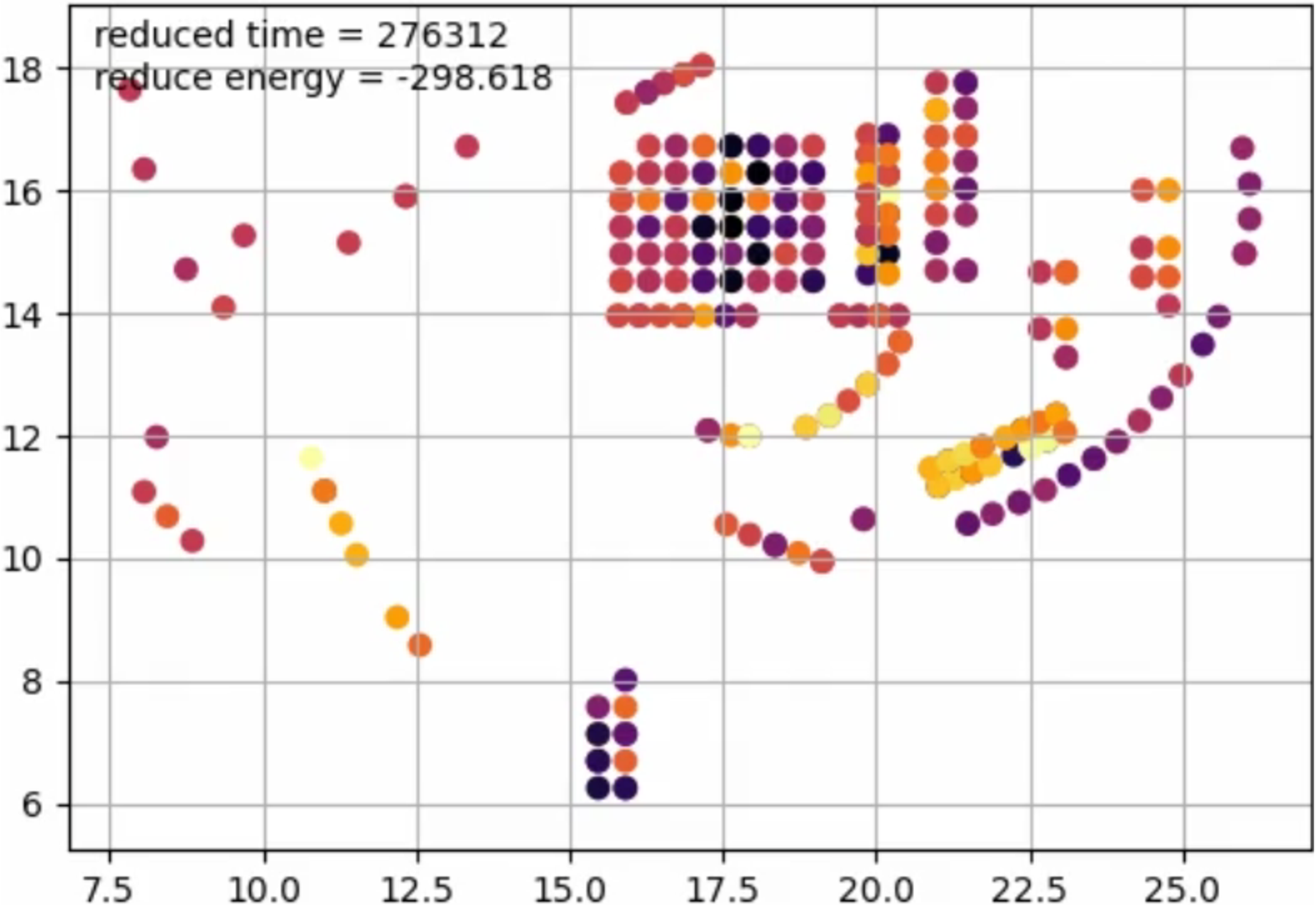}
	\caption{}
	\label{fig:animation12}
\end{subfigure}
\begin{subfigure}{.45\textwidth}
	\centering
	\includegraphics[width=1.0\linewidth]{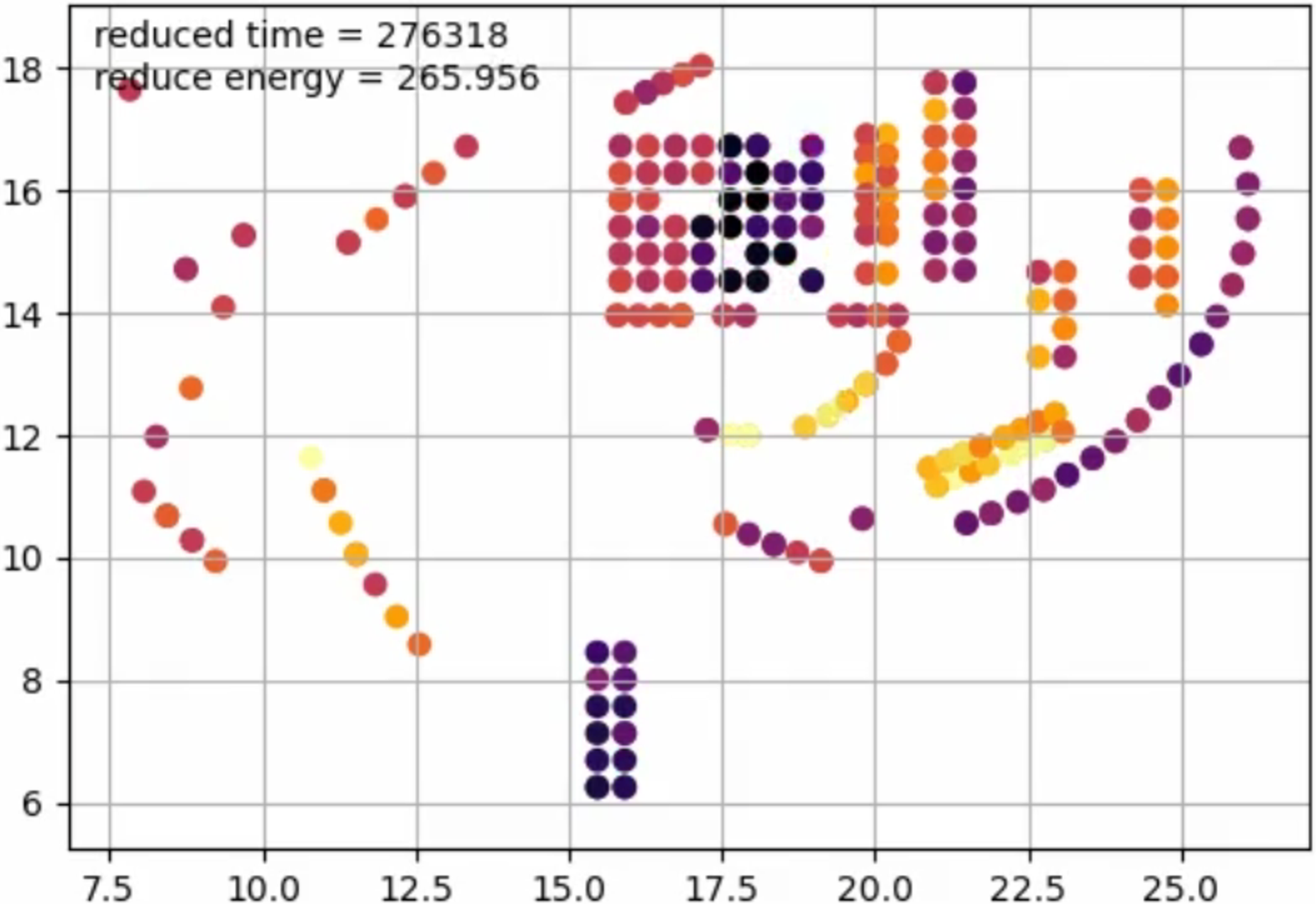}
	\caption{}
	\label{fig:animation13}
\end{subfigure}
\begin{subfigure}{.45\textwidth}
	\centering
	\includegraphics[width=1.0\linewidth]{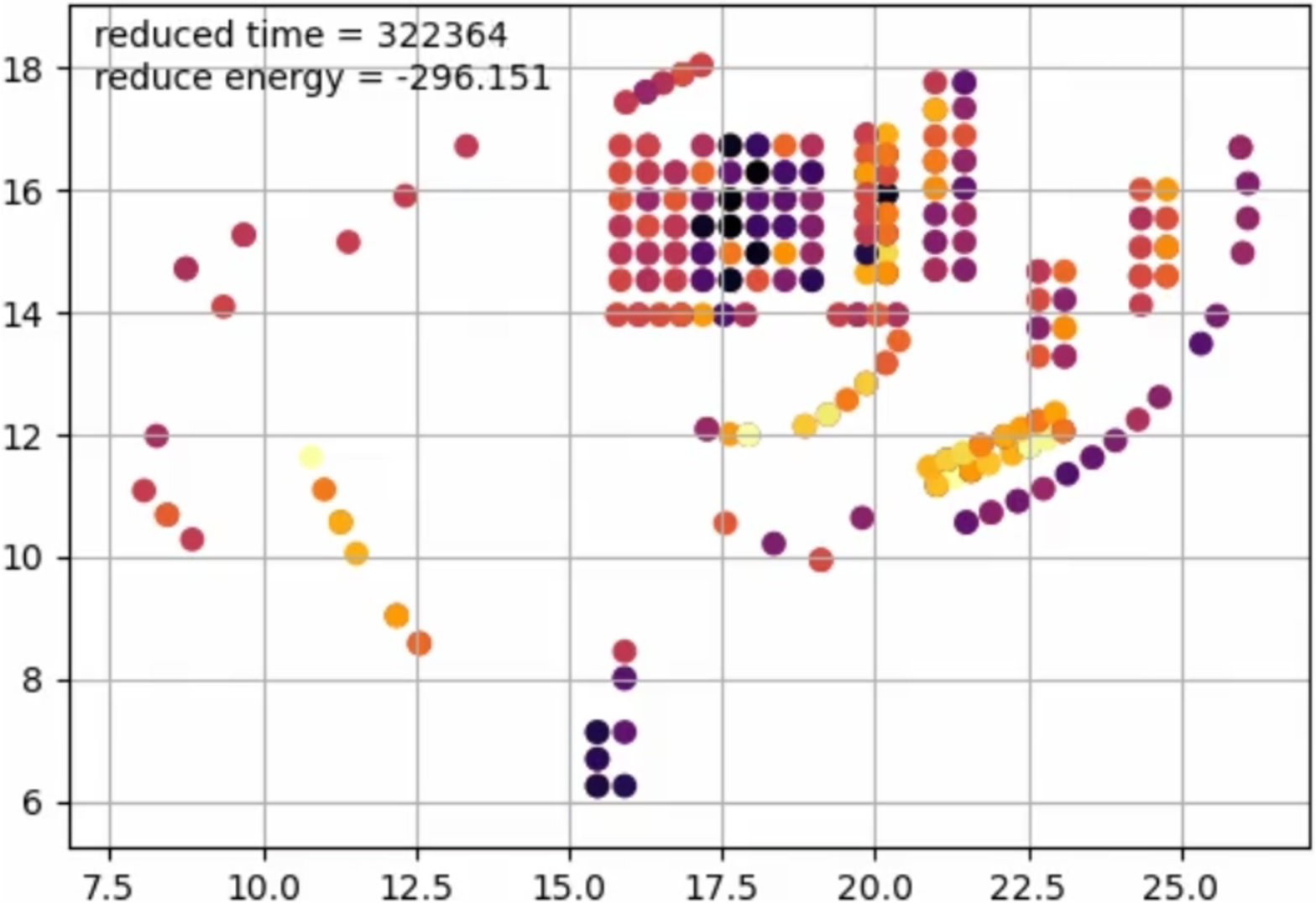}
	\caption{}
	\label{fig:animation14}
\end{subfigure}
\begin{subfigure}{.45\textwidth}
	\centering
	\includegraphics[width=1.0\linewidth]{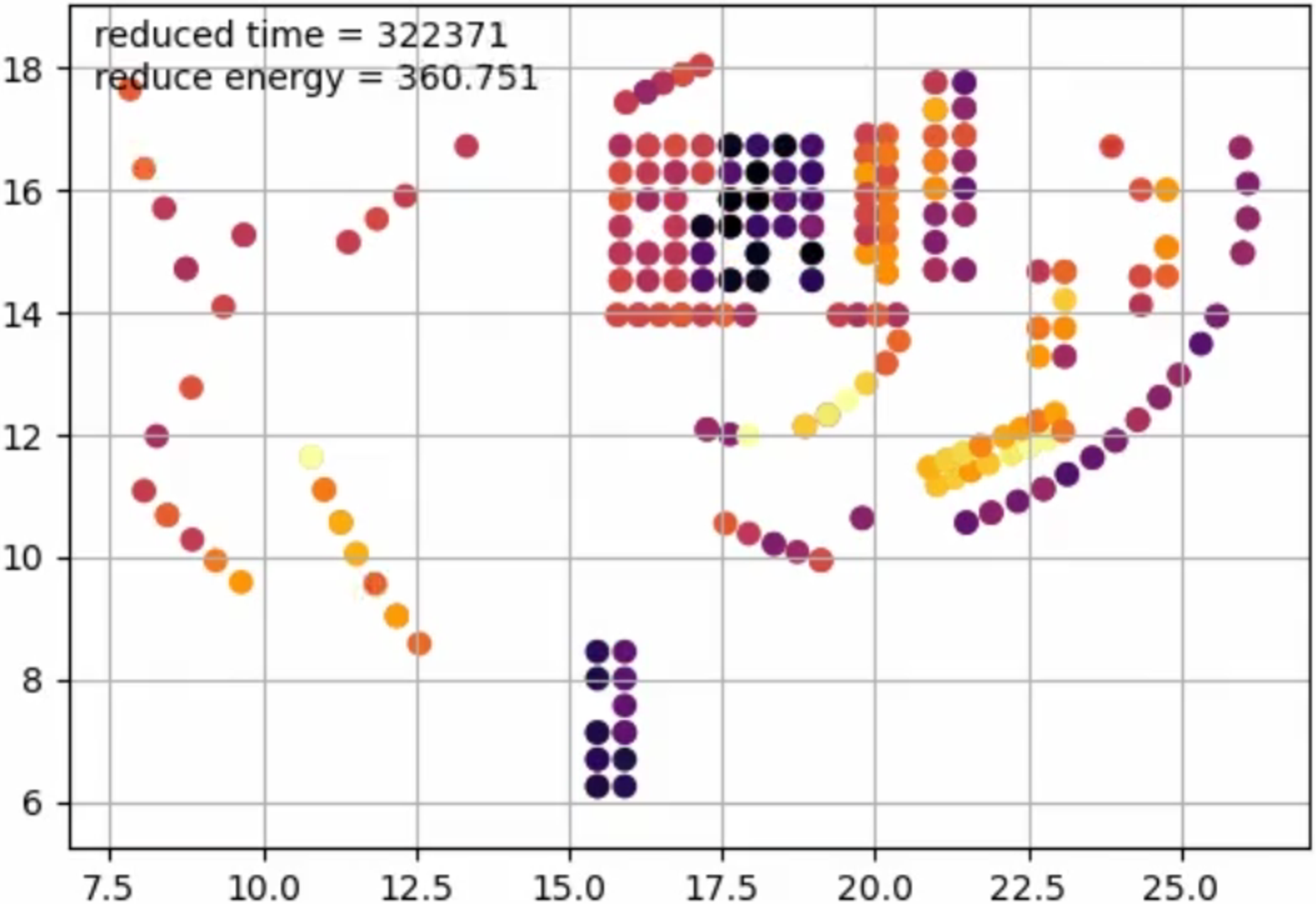}
	\caption{}
	\label{fig:animation15}
\end{subfigure}
		\caption{Animation snapshots of the featuring iterations with the highest and lowest energy in the eight periods in Figure $\ref{fig:energyVsIteration}$. Stands colors have the same meaning as the color bar in Fig.$\ref{fig:initLocsOfAtoms}$ }
		\label{fig:animations}
	\end{figure}
\end{center}
\newpage
In addition, we also plotted the energy gradient (zoomed in between range $\pm 15$) vs. iterations in Figure $\ref{fig:energyGradient}$. Inset was the energy gradient for the whole range in the vertical axis. Envelopes formed by the curves in every period were almost very similar. This may demonstrate that the initial configuration of the system may be irrelevant for the algorithm to search for a configuration with tolerable energy difference to the desired optimal energy value.
With the computer specs of Intel Core i7-4500U CPU @ 1.80 GHz 2.39 GHz and RAM 8GB, 64-bit operating system and x64-based processor, it took as about $5$ hours to minimize energy from the initial $477.55282$ to the optimal $-300.00073$. Codes may be accessed in Ref.\cite{macao food festival code V4}.
Figure $\ref{fig:optimalLocOfAtoms}$ showed the optional locations of stands. Compared with the initial locations of stands in Fig.$\ref{fig:initLocsOfAtoms}$, it had a significant difference. Generally speaking, the stands were well organized in the optimal layout as expected. Then, stands in Fig.$\ref{fig:optimalLocOfAtoms}$ were divided mainly into two parts apparently, mostly in the area of the x-axis $15.0$ to $22.5$, and the y-axis $14$ to $18$. One was that most of the stands were located on the upper right half the diagram, while lower left half part was composed of very few stands aligned with two columns, together with some sporadic booths surrounding around the Square holding the event.  One interesting feature was that there were $2$ parallel columns of stands, located along $x=20.0$ and the other skew one along near $y=12$, alternately consisting of booths with the higher $Q\prime$ and the lower ones. This interesting arrangement demonstrated that it may be a good idea to allocate together popular booths with less popular ones. On the other hand, very few of the stands were dispersed in the area from x-axis $7.5$ to $15.0$, and the y-axis $10$ to $18$. This layout may accomplish our goal of allocating stands as close as possible while minimizing the overall electrostatic potential energy. The complete list of stand locations were given in Table $\ref{tab:coordinates of optimal layout plan}$.
Electrostatic theorem was further brought upon with real-world applications once we interpret that the electrostatic energy density is the density of customers, while the electric field line points to the reversed net crowd flow at the particular location. Fig. $\ref{fig:electricFieldLine}$ showed the electric field vectors\cite{electric field lines} of the optimal layout plan within $1000\times1000 $ grid. Vectors indicated the reversed net crowd flows, with directions toward high $Q\prime$ while outward away from the lower $Q\prime$. Moreover, Fig. $\ref{fig:energyDensity}$ showed the logarithmic energy density, which was the prediction of customer density at the Festival. By calculating the energy density and electric field line, we may be able to predict the customer density and net movement, making it possible to enhance customer service at the festival.
\section{Conclusions}
	We successfully built up a mathematical model to calculate the optimal layout plan for stands at the Macau Food Festival while treating the popularity of stands as the Effective Charges. The popularity of every stand was acquired by implementing questionnaire surveys. Stands were purportedly treated as charged particles carrying some effective charge with the concept of more popular stands carrying higher EffQs. Also, to avoid customers from gathering together around popular stands, as well as the requirements that stands should not be too far away from one another, it is convincing that the optimal layout plan could be justified by searching for the global minimum of Coulomb energy for a certain configuration of stand locations. The above reasoning may serve as a guideline for future similar applications of this method. Since the landmark annual event Macau Food Festival enhances the local economy effectively and attracts a large number of tourists, our model may help prevent customers from congregating around more popular stands effectively.\\
	Our SAMA accomplished the purpose and reduced the initial energy from $477.55282$ to $-300.00073$ about $5$ hours (computer specs: Intel Core i7-4500U CPU @ $1.80$ GHz $2.39$ GHz and RAM 8GB, 64-bit operating system and x64-based processor). Moreover, we also proposed the optimal locations for booths in Figure $\ref{fig:optimalLocOfAtoms}$, with detailed coordinates listed in Table $\ref{tab:coordinates of optimal layout plan}$. Besides, electrostatic energy density is interpreted as density of customers, and electric field is the reversed crowd flow. Therefore, our model could be used to predict the density of customers and net crowd flow at any specific locations at the event, making it possible to assist a better tourist experience.\\	
	The effect of initial configuration of stands on the search of optimal energy value was further investigated by resetting the system several times until the desired energy value was found. We may observe that it could be significant for the system to be initialized with a proper configuration for SAMA to attain a global minimum, whereas computation time was irrelevant to the initial configuration if one only requested an approximate configuration energy close enough to the global minimum.\\	
	All in all, it is our pleasure that the model may be executed in future events to providing an optimization of the booth layout, stimulating consumption and boost economic development for Macau.
	\section{Acknowledgment}
	We thank Pui Ching Middle School in Macau PRC for the kindness to support this research project.

\newpage
\begin{appendices}
\begin{landscape}
	\section{Names, Coordinates and ID numbers of Stands}
\begin{longtable}{llllll}
		\caption{Names, Coordinates and ID numbers of Stands. Number J1 to J60 were $60$ empty locations that we chose by our own discretion in order to find out lower value of energy configuration.}
		\label{tab:Names, Coordinates and ID numbers of Stands}\\
		\hline
		standID & number & type                     & name                                                       & x       & y       \\ \hline
		1       & A1     & chinese restaurant zone  & shanghai 456 restaurant                                    & 25.9754 & 16.6909 \\
		2       & A2     & chinese restaurant zone  & metró pole                                                 & 26.0947 & 16.106  \\
		3       & A3     & chinese restaurant zone  & wong long frutos do mar casa de pasto                      & 26.0947 & 15.535  \\
		4       & A4     & chinese restaurant zone  & estabelecmento de comdas teng hou                          & 26.0125 & 14.9698 \\
		5       & A5     & chinese restaurant zone  & li jing xuan restaurante                                   & 25.8371 & 14.4576 \\
		6       & A6     & chinese restaurant zone  & restaurante beautiful world                                & 25.5923 & 13.9432 \\
\end{longtable}
\end{landscape}

\begin{sidewaystable}  
	\section{Grand questionnaire}
	\tabcolsep=0.26cm
	\centering
	\resizebox{\columnwidth}{!}{
\begin{tabular}{lccccccc}
	\hline
	\diagbox[width=11em]{Time stamp}{Type} & Chinese restaurant zone & european delicacies zone & Dessert zone & Asia delicicas zone & Local delicicas zone & Game booth & Sponsor and Other \\ \hline
	2021/11/5 am 10:58:33      & 9                       & 7                        & 10           & 9                   & 8                    & 8          & 6                 \\
	2021/11/5 am 10:59:29      & 4                       & 3                        & 8            & 6                   & 7                    & 9          & 8                 \\
	2021/11/5 am 11:00:00      & 7                       & 8                        & 9            & 8                   & 7                    & 9          & 7                 \\
	2021/11/5 am 11:01:45      & 9                       & 4                        & 10           & 7                   & 4                    & 10         & 8                 \\
	2021/11/5 am 11:06:10      & 10                      & 9                        & 9            & 9                   & 8                    & 3          & 5                 \\
	2021/11/8 am 8:28:54       & 10                      & 5                        & 9            & 9                   & 3                    & 3          & 3                 \\
	2021/11/8 am 8:54:37       & 5                       & 8                        & 8            & 9                   & 7                    & 4          & 4                 \\
\hline
\end{tabular}
}
	\caption{Grand questionnaire.}
	\label{tab:grand questionnaire}
\end{sidewaystable}

\begin{sidewaystable}  
	\section{Chinese restaurant zone}
	\tabcolsep=0.18cm
	\centering
\begin{tabular}{lccccccccccccccccc}
	\hline
	\diagbox[width=11em]{Time stamp}{StandID} & 1     & 2     & 3     & 4     & 5     & 6     & 7     & 8     & 9     & 10    & 11    & 12    & 13    & 14    & 15    & 16    & 17    \\ \hline
	2021/11/3 pm 7:47:26       & 6     & 4     & 5     & 5     & 4     & 3     & 5     & 7     & 7     & 6     & 4     & 4     & 4     & 5     & 4     & 5     & 6     \\
	2021/11/5 am 11:02:20      & 10    & 7     & 9     & 8     & 10    & 5     & 8     & 9     & 6     & 9     & 6     & 10    & 6     & 10    & 6     & 9     & 7     \\
	2021/11/5 am 11:02:46      & 9     & 7     & 9     & 8     & 7     & 10    & 9     & 10    & 7     & 9     & 7     & 8     & 9     & 8     & 7     & 10    & 10    \\
	2021/11/5 am 11:05:48      & 7     & 6     & 6     & 5     & 4     & 5     & 5     & 5     & 5     & 3     & 8     & 6     & 6     & 6     & 5     & 5     & 8     \\
	2021/11/5 am 11:06:42      & 5     & 10    & 3     & 8     & 4     & 6     & 8     & 3     & 7     & 6     & 4     & 6     & 4     & 4     & 8     & 4     & 7     \\
	2021/11/5 am 11:09:39      & 2     & 4     & 10    & 10    & 8     & 5     & 9     & 9     & 4     & 3     & 2     & 10    & 10    & 10    & 2     & 10    & 3     \\
	2021/11/17 pm 11:22:45     & 7     & 4     & 5     & 5     & 5     & 1     & 6     & 3     & 7     & 5     & 6     & 2     & 1     & 5     & 5     & 7     & 5     \\
\hline
\end{tabular}
	\caption{Chinese restaurant zone.}
	\label{tab:Chinese restaurant zone}
\end{sidewaystable}

\begin{sidewaystable}  
	\section{European delicacies zone}
	\tabcolsep=0.18cm
	\centering
\begin{tabular}{lccccccccccc}
	\hline
	\diagbox[width=11em]{Time stamp}{StandID} & 18    & 19    & 20    & 21    & 22    & 23    & 24    & 25    & 26    & 27    & 28    \\ \hline
	2021/11/5 am 10:59:06  & 8     & 10    & 5     & 10    & 10    & 4     & 9     & 7     & 10    & 7     & 10    \\
	2021/11/5 am 11:01:03  & 7     & 8     & 7     & 8     & 7     & 7     & 5     & 6     & 5     & 6     & 6     \\
	2021/11/5 am 11:01:39  & 8     & 5     & 7     & 5     & 4     & 8     & 7     & 5     & 4     & 8     & 9     \\
	2021/11/5 am 11:02:21  & 8     & 3     & 3     & 7     & 4     & 5     & 9     & 10    & 4     & 6     & 8     \\
	2021/11/5 am 11:08:33  & 9     & 10    & 5     & 3     & 3     & 9     & 9     & 9     & 4     & 10    & 3     \\
	2021/11/8 am 8:29:15   & 8     & 3     & 5     & 7     & 4     & 3     & 3     & 9     & 7     & 9     & 6     \\
	2021/11/8 pm 12:29:31  & 7     & 7     & 6     & 6     & 7     & 7     & 7     & 7     & 6     & 8     & 6     \\
\hline
\end{tabular}
	\caption{European delicacies zone.}
	\label{tab:European delicacies zone}
\end{sidewaystable}

\begin{sidewaystable}  
	\section{Dessert zone}
	\tabcolsep=0.18cm
	\resizebox{\columnwidth}{!}{
\begin{tabular}{lcccccccccccccccccccccccccccc}
	\hline
	\diagbox[width=11em]{Time stamp}{StandID} & 29    & 30    & 31    & 32    & 33    & 34    & 35    & 36    & 37    & 38    & 39    & 40    & 41    & 42    & 43    & 44    & 45    & 46    & 47    & 48    & 49    & 50    & 51    & 52    & 53    & 54    & 55    & 56    \\ \hline
	2021/11/5 am 11:00:03  & 10    & 9     & 8     & 10    & 8     & 10    & 9     & 8     & 10    & 8     & 7     & 9     & 9     & 9     & 7     & 8     & 9     & 8     & 9     & 8     & 7     & 8     & 9     & 7     & 9     & 8     & 8     & 8     \\
	2021/11/5 am 11:02:51  & 9     & 7     & 10    & 9     & 10    & 9     & 9     & 10    & 10    & 9     & 9     & 9     & 4     & 6     & 2     & 4     & 4     & 7     & 7     & 6     & 5     & 5     & 7     & 8     & 7     & 6     & 9     & 8     \\
	2021/11/5 am 11:02:58  & 9     & 4     & 6     & 7     & 8     & 6     & 6     & 9     & 7     & 6     & 8     & 8     & 6     & 5     & 7     & 8     & 7     & 6     & 8     & 5     & 4     & 7     & 8     & 6     & 5     & 7     & 6     & 4     \\
	2021/11/5 am 11:04:55  & 9     & 3     & 2     & 8     & 8     & 3     & 8     & 3     & 9     & 3     & 9     & 3     & 3     & 9     & 10    & 2     & 9     & 9     & 2     & 7     & 1     & 1     & 4     & 10    & 1     & 5     & 7     & 1     \\
	2021/11/8 pm 12:31:39  & 8     & 8     & 8     & 8     & 8     & 8     & 9     & 7     & 8     & 8     & 9     & 9     & 8     & 7     & 7     & 8     & 8     & 9     & 7     & 8     & 8     & 8     & 8     & 9     & 8     & 9     & 8     & 8     \\
	2021/11/17 pm 11:19:44 & 7     & 6     & 8     & 6     & 8     & 6     & 8     & 6     & 7     & 8     & 5     & 5     & 5     & 7     & 4     & 7     & 4     & 3     & 4     & 7     & 6     & 8     & 6     & 7     & 5     & 4     & 4     & 4     \\
\hline
\end{tabular}
	}
	\caption{Dessert zone.}
	\label{tab:Dessert zone}
\end{sidewaystable}

\begin{sidewaystable}  
	\section{Asia delicicas zone}
	\tabcolsep=0.18cm
	\resizebox{\columnwidth}{!}{
\begin{tabular}{lccccccccccccccccccccccccc}
	\hline
	\diagbox[width=11em]{Time stamp}{StandID} & 57    & 58    & 59    & 60    & 61    & 62    & 63    & 64    & 65    & 66    & 67    & 68    & 69    & 70    & 71    & 72    & 73    & 74    & 75    & 76    & 77    & 78    & 79    & 80    & 81    \\ \hline
	2021/11/3 pm 7:48:33   & 7     & 6     & 9     & 6     & 5     & 5     & 4     & 4     & 4     & 4     & 8     & 6     & 4     & 6     & 6     & 5     & 4     & 4     & 4     & 5     & 4     & 6     & 6     & 4     & 6     \\
	2021/11/5 am 11:01:56  & 10    & 8     & 9     & 10    & 9     & 8     & 10    & 7     & 10    & 10    & 7     & 10    & 9     & 7     & 9     & 9     & 8     & 7     & 10    & 8     & 10    & 7     & 10    & 6     & 10    \\
	2021/11/5 am 11:05:17  & 9     & 9     & 7     & 7     & 7     & 6     & 7     & 7     & 7     & 5     & 6     & 5     & 7     & 9     & 10    & 9     & 8     & 7     & 8     & 7     & 6     & 8     & 8     & 6     & 7     \\
	2021/11/5 am 11:06:21  & 5     & 4     & 7     & 6     & 3     & 8     & 5     & 8     & 4     & 8     & 4     & 4     & 7     & 5     & 6     & 8     & 10    & 10    & 4     & 6     & 8     & 7     & 9     & 9     & 9     \\
	2021/11/5 am 11:08:39  & 9     & 3     & 2     & 5     & 3     & 8     & 9     & 2     & 9     & 4     & 9     & 10    & 10    & 2     & 9     & 4     & 10    & 4     & 2     & 2     & 10    & 2     & 1     & 1     & 10    \\
	2021/11/8 pm 12:33:34  & 5     & 5     & 7     & 6     & 7     & 6     & 7     & 7     & 7     & 7     & 8     & 6     & 9     & 9     & 9     & 9     & 9     & 9     & 8     & 9     & 9     & 8     & 9     & 9     & 8     \\
	2021/11/17 pm 11:21:58 & 6     & 6     & 5     & 6     & 6     & 6     & 5     & 8     & 7     & 8     & 7     & 6     & 6     & 4     & 6     & 6     & 7     & 7     & 7     & 6     & 6     & 7     & 6     & 6     & 4     \\
\hline
\end{tabular}
	}
	\caption{Asia delicicas zone.}
	\label{tab:Asia delicicas zone}
\end{sidewaystable}

\begin{sidewaystable}  
	\section{Local delicicas zone}
	\tabcolsep=0.18cm
	\resizebox{\columnwidth}{!}{
\begin{tabular}{lcccccccccccccccccccccccccccccccccccc}
	\hline
	\diagbox[width=11em]{Time stamp}{StandID} & 82    & 83    & 84    & 85    & 86    & 87    & 88    & 89    & 90    & 91    & 92    & 93    & 94    & 95    & 96    & 97    & 98    & 99    & 100   & 101   & 102   & 103   & 104   & 105   & 106   & 107   & 108   & 109   & 110   & 111   & 112   & 113   & 114   & 115   & 116   & 117   \\ \hline
	2021/11/3 pm 7:50:48   & 8     & 7     & 8     & 6     & 6     & 5     & 6     & 6     & 4     & 7     & 9     & 8     & 7     & 6     & 7     & 7     & 5     & 4     & 4     & 7     & 5     & 6     & 7     & 7     & 7     & 4     & 7     & 5     & 5     & 4     & 5     & 3     & 4     & 2     & 3     & 6     \\
	2021/11/5 am 11:04:06  & 9     & 8     & 7     & 7     & 6     & 7     & 6     & 9     & 8     & 9     & 10    & 9     & 8     & 8     & 7     & 7     & 8     & 5     & 6     & 5     & 6     & 5     & 5     & 8     & 8     & 6     & 8     & 8     & 6     & 7     & 6     & 8     & 7     & 8     & 7     & 6     \\
	2021/11/5 am 11:05:17  & 10    & 1     & 7     & 3     & 4     & 2     & 7     & 6     & 10    & 10    & 10    & 8     & 5     & 10    & 2     & 2     & 5     & 1     & 3     & 4     & 5     & 7     & 5     & 4     & 6     & 4     & 8     & 4     & 8     & 4     & 6     & 5     & 10    & 10    & 6     & 8     \\
	2021/11/5 am 11:06:27  & 9     & 6     & 7     & 1     & 2     & 3     & 2     & 8     & 7     & 9     & 10    & 3     & 10    & 3     & 6     & 4     & 5     & 5     & 4     & 3     & 3     & 2     & 2     & 3     & 1     & 1     & 1     & 10    & 1     & 2     & 4     & 9     & 9     & 2     & 6     & 4     \\
	2021/11/8 am 8:56:59   & 7     & 7     & 5     & 6     & 6     & 5     & 5     & 5     & 6     & 9     & 9     & 8     & 8     & 8     & 7     & 7     & 7     & 8     & 7     & 8     & 8     & 7     & 7     & 7     & 6     & 7     & 5     & 7     & 6     & 6     & 6     & 6     & 8     & 10    & 6     & 2     \\
\hline
\end{tabular}
	}
	\caption{Local delicicas zone.}
	\label{tab:Local delicicas zone}
\end{sidewaystable}

\begin{sidewaystable}  
	\section{Sponsor and other}
	\tabcolsep=0.18cm
	\centering
\begin{tabular}{lcccccccccccc}
	\hline
	\diagbox[width=11em]{Time stamp}{StandID} & 118   & 119   & 120   & 121   & 122   & 123   & 124   & 125   & 137   & 138   & 139   & 140   \\ \hline
	2021/11/3 pm 7:49:23   & 3     & 4     & 4     & 3     & 4     & 7     & 5     & 6     & 6     & 5     & 8     & 8     \\
	2021/11/5 am 11:01:25  & 10    & 8     & 8     & 7     & 9     & 7     & 10    & 8     & 9     & 9     & 7     & 9     \\
	2021/11/5 am 11:04:33  & 7     & 8     & 3     & 4     & 4     & 4     & 5     & 4     & 5     & 5     & 7     & 7     \\
	2021/11/5 am 11:05:40  & 6     & 4     & 7     & 3     & 8     & 4     & 6     & 7     & 4     & 5     & 7     & 6     \\
	2021/11/5 am 11:07:21  & 3     & 7     & 9     & 1     & 1     & 3     & 3     & 1     & 8     & 8     & 9     & 10    \\
	2021/11/5 am 11:07:36  & 3     & 7     & 6     & 4     & 7     & 3     & 9     & 6     & 7     & 4     & 2     & 7     \\
	2021/11/17 pm 11:21:17 & 3     & 3     & 2     & 6     & 4     & 4     & 4     & 4     & 7     & 7     & 6     & 5     \\
\hline
\end{tabular}
	\caption{Sponsor (stand ID: 118-125) and other (stand ID: 137-140).}
	\label{tab:Sponsor and other}
\end{sidewaystable}

\begin{sidewaystable}  
	\section{Game booth}
	\tabcolsep=0.2cm
	\centering
\begin{tabular}{lccccccccccc}
	\hline
	\diagbox[width=11em]{Time stamp}{StandID} & 126   & 127   & 128   & 129   & 130   & 131   & 132   & 133   & 134   & 135   & 136   \\ \hline
	2021/11/3 pm 7:45:29   & 7     & 5     & 6     & 4     & 4     & 6     & 8     & 9     & 6     & 4     & 7     \\
	2021/11/5 am 10:53:36  & 9     & 4     & 7     & 5     & 8     & 7     & 6     & 10    & 8     & 6     & 7     \\
	2021/11/5 am 10:59:25  & 10    & 9     & 9     & 7     & 7     & 9     & 7     & 10    & 9     & 8     & 7     \\
	2021/11/5 am 11:01:41  & 7     & 6     & 8     & 8     & 9     & 9     & 10    & 10    & 9     & 9     & 7     \\
	2021/11/5 am 11:03:29  & 10    & 3     & 3     & 9     & 9     & 10    & 10    & 3     & 9     & 5     & 2     \\
	2021/11/5 am 11:10:18  & 3     & 2     & 4     & 10    & 2     & 3     & 3     & 3     & 2     & 3     & 2     \\
	2021/11/8 pm 12:38:14  & 8     & 6     & 7     & 8     & 6     & 7     & 8     & 8     & 8     & 7     & 6     \\
\hline
\end{tabular}
	\caption{Game booth.}
	\label{tab:Game booth}
\end{sidewaystable}

\clearpage 
\section{Normalized effective charges $Q\prime$ of stands}
	\begin{longtable}{cccccccc}
		\caption{The normalized effective charges $Q\prime$ of stands were obtained as follows. First, we calculated $\widebar{x}$ and $\sigma_{\widebar{x}}$, referring to the mean of the popularity of the sub-categorical stand, and the standard deviation of mean for the particular stand, respectively. Second, we also calculated $\widebar{y}$ and $\sigma_{\widebar{y}}$, respectively referring to the mean of the popularity from the grand questionnaire for a specific type of stands, and the standard deviation of mean for that particular type of stands. Third, we calculated the final mean of each stand by calculating $\widebar{xy}=\widebar{x}\widebar{y}$, together with the error propagation $\sigma_{\widebar{xy}}$ described in Eq.($\ref{error propagatoin}$). Finally we also calculated $Q\prime$s by Eq.($\ref{standarScaler}$) at the last column, which were the effective charges of stands in our algorithm.}
		\label{tab:norm Q}\\
		\hline
StandID & Mean($\widebar{x}$)  & SD of mean($\sigma_{\widebar{x}}$)& \begin{tabular}[c]{@{}c@{}}Mean($\widebar{y}$) from\\ Grand Questionnaire\end{tabular} & \begin{tabular}[c]{@{}c@{}}sd of mean($\sigma_{\widebar{y}}$) from \\ Grand Questionnaire\end{tabular} & \begin{tabular}[c]{@{}c@{}}Final mean\\($\widebar{xy}$) \end{tabular}& \begin{tabular}[c]{@{}c@{}}Final sd of\\ mean($\sigma_{\widebar{xy}}$)\end{tabular} & $Q\prime$ \\ \hline
1       & 6.875 & 2.459      & 6.543                                                                   & 0.403                                                                          & 44.983     & 16.328           & -0.443  \\
2       & 6.594 & 2.354      & 6.543                                                                   & 0.403                                                                          & 43.143     & 15.627           & -0.696  \\
\hline
	\end{longtable}

\section{Coordinates of stands for the initial layout plan}
\begin{longtable}{cccc}
	\caption{Coordinates of stands for the initial layout plan. Integer $-1$ in Columns \textbf{StandID} and \textbf{Charge} actually means that there is \textbf{NO} stand at that particular location.}
	\label{tab:coordinates of initial layout plan}\\
	\hline
standID & X & Y & charge \\ \hline
0 & 25.9754 & 16.6909 & -0.444 \\
1 & 26.0947 & 16.106 & -0.698 \\
2 & 26.0947 & 15.535 & -0.501 \\
3 & 26.0125 & 14.9698 & -0.416 \\
-1 & 25.8371 & 14.4576 & -1.000 \\
4 & 25.5923 & 13.9432 & -0.754 \\
\hline
\end{longtable}

\section{Coordinates of stands for the optimal layout plan}
\begin{longtable}{cccc}
	\caption{Coordinates of stands for the optimal layout plan. Integer $-1$ in Columns \textbf{StandID} and \textbf{Charge} actually means that there is \textbf{NO} stand at that particular location.}
	\label{tab:coordinates of optimal layout plan}\\
	\hline
standID & X & Y & charge \\ \hline
-1 & 25.9754 & 16.6909 & -1 \\
-1 & 26.0947 & 16.106 & -1 \\
23 & 26.0947 & 15.535 & 0.009 \\
86 & 26.0125 & 14.9698 & 0.007 \\
-1 & 25.8371 & 14.4576 & -1 \\
-1 & 25.5923 & 13.9432 & -1 \\
\hline
\end{longtable}
\end{appendices}
\end{document}